\documentclass{article}
\usepackage{import}
\usepackage{geometry}
\usepackage[utf8]{inputenc}
\usepackage[T1]{fontenc}
\usepackage{textcomp}
\usepackage{textgreek}
\usepackage[american]{babel}
\usepackage[autostyle,autopunct=true]{csquotes}
\usepackage[
    labelfont={bf},
    labelsep=period,
    font={small,sf},
    singlelinecheck=off,
    tableposition=top
]{caption}
\usepackage{subcaption}
\usepackage[font={small,sf}]{floatrow}
\usepackage{multirow}
\usepackage{booktabs}
\usepackage{threeparttable}
\usepackage[square,sort,comma,numbers]{natbib}
\usepackage{graphicx}
\usepackage{tcolorbox}
\usepackage{enumitem}
\usepackage[
    colorlinks=true,
    linkcolor=blue,
    citecolor=blue,
    urlcolor=blue,
    pdfencoding=auto,
    psdextra
]{hyperref}
\usepackage{amsmath}
\usepackage{amssymb}
\usepackage{amsthm}
\usepackage{mathtools}
\usepackage{mathbbol}
\usepackage{titlesec}
\usepackage{placeins}
\usepackage[title]{appendix}
\usepackage{tikz}
\usetikzlibrary{%
    shapes,
    shapes.geometric,
    positioning,
    shadows,
    trees,
    calc,
    arrows,
    fit
}

\pgfdeclarelayer{background}
\pgfdeclarelayer{foreground}
\pgfsetlayers{background,main,foreground}

\definecolor{myred}{HTML}{EB7159}
\definecolor{myblue}{HTML}{66C8E6}

\definecolor{c1}{HTML}{EB7159}
\definecolor{c2}{HTML}{66C8E6}
\definecolor{c3}{HTML}{AB65B5}
\definecolor{c4}{HTML}{829F49}
\definecolor{c5}{HTML}{6876CE}
\definecolor{c6}{HTML}{BE8A3B}
\definecolor{c7}{HTML}{CF5786}
\definecolor{c8}{HTML}{AD483A}

\tikzset{
    node distance=1cm,
    basic/.style = {
        draw,
        text width=1cm,
        drop shadow,
        font=\sffamily,
        rectangle,
        fill=lightgray!30,
        align=center
    },
    plain/.style = {
        draw=none
    },
    label/.style = {
        font=\bfseries\sffamily,
        align=center
    },
    slabel/.style = {
        label,
        font = \bfseries\sffamily\small
    },
    lab/.style = {
        label,
        font=\sffamily
    },
    tlab/.style = {
        font=\sffamily\small
    },
    main node/.style = {
        circle,
        fill=blue!20,
        draw,
        minimum size=1cm,
        inner sep=0pt
    },
    vertex/.style = {
        draw,
        circle,
        drop shadow,
        font=\sffamily\boldmath,
        align=center,
        color=black,
        fill=black!10
    },
    ego/.style = {
        vertex,
        color=black,
        fill=myred
    },
    focal/.style = {
        ego
    },
    sim/.style = {
        ego,
    },
    comp/.style = {
        ego,
        fill=myblue
    },
    mix/.style = {
        ego,
        fill=violet!70
    },
    focus/.style = {
        ego,
        color=black,
        fill=yellow
    },
    edge/.style = {
        ultra thick
    },
    dedge/.style = {
        edge,
        -latex
    },
    weight/.style = {
        label,
        circle,
        fill=white,
        draw,
        minimum size=3mm,
        inner sep=1pt
    },
    elabel/.style = {
        midway,
        sloped
    },
    plus/.style = {
        elabel,
        color=myblue
    },
    minus/.style = {
        elabel,
        color=red!60
    },
    circ/.style = {
        draw,
        shape=circle
    }
}

\titleformat{\section}%
    {\normalfont\Large\bfseries}{\thesection.}{1em}{}
\titleformat{\subsection}%
    {\normalfont\large\bfseries}{\thesubsection.}{1em}{}
\titleformat{\subsubsection}%
    {\normalfont\normalsize\bfseries}{\thesubsubsection.}{1em}{}

\DeclareMathOperator*{\tr}{tr}

\newcommand{\mean}[1]{\ensuremath{\langle#1\rangle}}

\newcommand{\norm}[1]{\left\lVert#1\right\rVert}

\newcommand{\rev}[1]{{#1}}
\newcommand{\revII}[1]{#1}
\usepackage{appendix}
\usepackage{lineno}

\graphicspath{{./figs/}}

\title{Polarization and multiscale structural balance in signed networks}
\author{
    Szymon Talaga$^{1\ast}$
    \and Massimo Stella$^{2}$
    \and Trevor James Swanson$^{3}$
    \and Andreia Sofia Teixeira$^{4}$
}
\date{
    {\normalsize
    $^1{}$Robert Zajonc Institute for Social Studies, University of Warsaw,\\
    Stawki 5/7, 00-183 Warsaw, Poland. \\
    $^2{}$Dipartimento di Psicologia e Scienze Cognitive, University of Trento\\
    Corso Bettini 84, 38068, Rovereto (TN), Italy \\
    $^3{}$Department of Psychology, University of Kansas, \\
    1415 Jayhawk Blvd, Lawrence, KS, 66045, USA \\
    $^4{}$LASIGE and Departamento de Informática, Faculdade de Ciências, Universidade de Lisboa, Lisboa, Portugal\\
    $^\ast{}$Corresponding author; E-mail: \texttt{stalaga@uw.edu.pl}
    }
}

\newtheorem{theorem}{Theorem}
\newtheorem*{corollary}{Corollary}

\newtheorem{definition}{Definition}

\begin{document}

\flushbottom
\maketitle
\thispagestyle{empty}

\begin{abstract}
\noindent
Polarization, understood as a division into mutually hostile groups, is a common feature of social systems.
It is studied in Structural Balance Theory (SBT) in terms of semicycles in signed
networks. However, enumerating semicycles is computationally expensive, so approximations 
are often needed.
Here we introduce Multiscale Semiwalk Balance (MSB) approach 
for measuring degree of balance (DoB) in (un)directed, (un)weighted signed networks by approximating semicycles with closed semiwalks. It allows for selection of the resolution of analysis
appropriate for assessing DoB motivated by Locality Principle (LP), which posits that patterns in shorter cycles are more important than in longer ones.
Our approach overcomes several limitations affecting walk-based approximations,
and provides methods for assessing DoB at various scales, from graphs to individual nodes,
and for clustering signed networks.
\revII{
We demonstrate its effectiveness by applying it to real-world social systems,
for which it produces explainable results consistent with expectations based on domain-specific knowledge.
}
\end{abstract}
\section{Introduction}

Networks are used in many branches of science and engineering for modeling complex systems.
Depending on the context, they may be undirected (ties are bidirectional)
or directed and weighted (ties have weights which usually indicate strength)
or unweighted~\cite{newmanNetworks2018}.
\revII{Moreover, some networks are signed, or have links that are either positive or negative,} 
and thus can be used to model valenced relations such as liking and disliking, or alliances and
war~\cite{doreianPartitioningApproachStructural1996,teixeiraEmergenceSocialBalance2017,estradaWalkbasedMeasureBalance2014,kirkleyBalanceSignedNetworks2019}. 
Signed networks are commonly used for representing systems capable of polarization, or clustering into 
groups with positive in-group and negative out-group ties. As a result, they have long been
important to social scientists interested in polarization and differentiation processes
inherent to formation of groups, attitudes and 
opinions~\cite{wassermanSocialNetworkAnalysis1994,doreianPartitioningApproachStructural1996,denooyLiteraryPlaygroundLiterary1999,schweighoferWeightedBalanceModel2020,nealSignTimesWeak2020}.
However, signed networks are also used in other disciplines for modeling diverse phenomena such as brain activation~\cite{saberiTopologicalImpactNegative2021}, ecological interactions~\cite{saizEvidenceStructuralBalance2017}, and financial 
time series~\cite{ferreiraLossStructuralBalance2021}. Moreover, it is often not only the signs that matter, but also the weights indicating intensities
of particular relations. Therefore, principled methods for analyzing signed networks, 
possibly with weights, are important for many applications.

\begin{figure}[htb!]
\centering
\includegraphics[width=.9\textwidth]{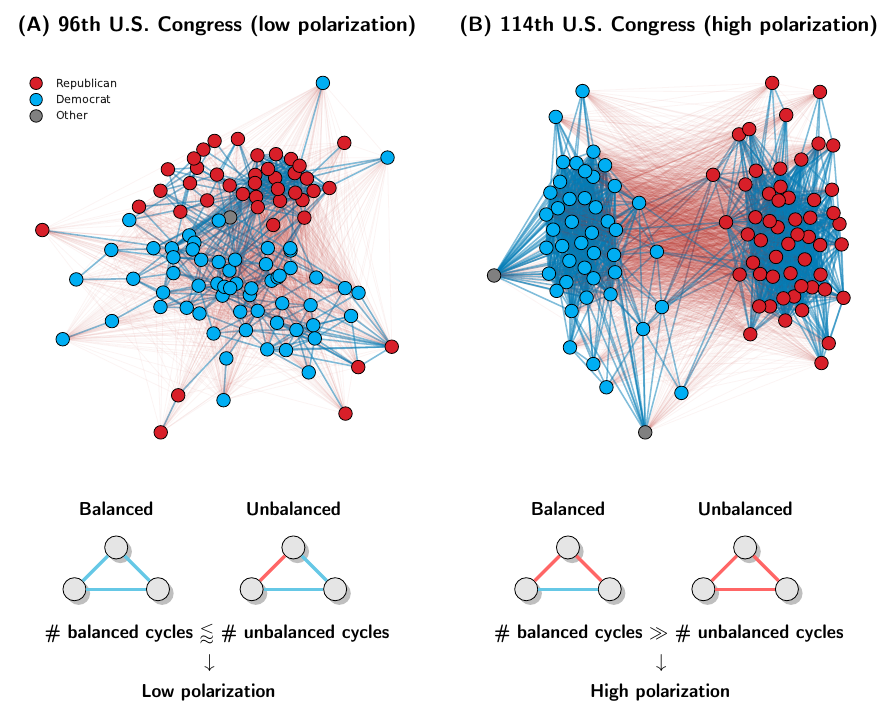}
\caption{
    Schematic explanation of the meaning of high and low polarization and
    its general connection to the frequencies of (strongly) balanced and unbalanced cycles.
    \textbf{(Top)} Two networks of bill co-sponsorship in the U.S. Senate. 
    Positive ties (blue) link senators, who tended to promote the same bills together 
    more often than by chance, and negative ties (red) correspond to those who collaborated
    less often than at random 
    (see Sec.~\ref{sec:results:congress} for the data description and a detailed analysis).
    \textbf{(Bottom)} All possible (strongly) balanced and unbalanced triads (3-cycles) and
    \revII{
        the general relationship between counts of (un)balanced cycles and polarization
        (note that, even though only 3-cycles are depicted, the general relationship pertains to
        cycles of all lengths). Low polarization is characterized by comparable frequencies,
        or more rarely by a majority of unbalanced cycles.
        High polarization implies that there is a clear majority of balanced cycles.
    }
    \textbf{(A)}~96th Congress (1979-81, Carter administration) was a period of low polarization
    with frequent positive between-party and negative within-party links. 
    This translated to comparable frequencies of balanced (positive product of edge signs)
    and unbalanced (negative product of edge signs) cycles.
    \textbf{(B)}~114th Congress (2015-17, Obama administration) featured high polarization
    with in-group ties being almost exclusively positive and out-group ties negative.
    This induced a distribution skewed towards balanced (positive) cycles.
}
\label{fig:examples}
\end{figure}

Since signed networks represent valenced relations, a fundamental question concerns 
the degree to which positive and negative ties are consistent with respect to notions
of (anti)transitivity, and whether these microscopic patterns give rise to a polarized macroscopic
organization into mutually antagonistic clusters. 
Both problems are studied in Structural Balance Theory (SBT)~\cite{cartwrightStructuralBalanceGeneralization1956,hararyStructuralModelsIntroduction1965},
which originated from Gestalt psychology and the work of Fritz Heider~\cite{heiderAttitudesCognitiveOrganization1946}, 
who proposed that positive relations should be transitive (a friend of my friend is my friend)
and negative relations antitransitive (an enemy of my enemy is my friend), 
\rev{
e.g.~two positively (negatively) linked nodes should have identical (opposite) signs on their ties
to shared neighbours.
}
These considerations were later formalized and generalized in graph-theoretic terms and used to
demonstrate that (anti)transitivity of (negative) positive relations is directly linked to the properties 
of cycles, and as a result to clustering and polarization. Namely, polarized 
\rev{
systems clustered in exactly two antagonistic groups,
in which in-group ties are exclusively positive and out-group ties negative,
require that all cycles are positive, 
or that the products of the signs of their edges are 
positive~\cite{cartwrightStructuralBalanceGeneralization1956}
(strong balance property; see Fig.~\ref{fig:examples} for a visual explanation and some examples).
}
Systems clustered into $b \geq 2$ antagonistic blocks require that there are no cycles with exactly
one negative edge (weak balance property)~\cite{davisClusteringStructuralBalance1967}. 
\rev{
See Methods, Sec.~\ref{sec:methods:sbt}, for an overview of the main definitions and theorems of SBT,
including their general form applicable to directed networks based on the notion of semicycles.
}

SBT specifies strict requirements for signed networks to be balanced 
(partitioned in antagonistic groups), but real-world systems are rarely organized
neatly enough to satisfy them completely. This is why a lot of work in SBT is concerned with measures
of the degree of balance (DoB), or partial balance~\cite{arefMeasuringPartialBalance2018}, 
which can be seen as indicators of a \enquote{distance} from the perfectly balanced state.
Such measures are typically directly or indirectly related to the relative frequencies of positive
and negative cycles (or cycles with exactly one negative edge in the case of weak balance).

\rev{
However, measuring structural balance in practice is not trivial. While defining DoB at the
level of cycles of a particular length $k$ is simple, since in this case the raw proportion
of balanced cycles is meaningful, any global DoB measure has to integrate information across
cycles of many different lengths and it is not immediately clear how this should be done.
The difficulty comes from the fact that typically longer cycles will be much more numerous
than shorter ones, so a simple proportion will be determined primarily by patterns found in long
cycles, but this may not be a desirable property. Indeed, already Cartwright and Harrary
hypothesized that shorter cycles should matter more when evaluating 
DoB~\cite{cartwrightStructuralBalanceGeneralization1956}. 
Moreover, this intuition has been later justified empirically by demonstrating that it is easier 
for people to memorize valences of ties in shorter cycles~\cite{zajoncStructuralBalanceReciprocity1965}.
More recently, analyses based on counting simple cycles demonstrated that real networks often 
have a relatively low cycle length threshold after which DoB measures quickly decrease, 
indicating that structural balance is found primarily in structures at smaller
scales~\cite{giscardEvaluatingBalanceSocial2017}.
}

Applying SBT in practice is \rev{further} complicated by the fact that enumerating and counting 
cycles is computationally expensive, especially for large graphs. 
\rev{
This problem can be partially alleviated with novel algorithms and sampling
methods, but exact solutions will always remain prohibitively expensive due to the nature of the problem.
Moreover, the current state-of-the-art sampling methods~\cite{giscardEvaluatingBalanceSocial2017}
are limited to \enquote{grayscale} measures which quantify DoB for cycles of particular lengths
and they do not offer any principled way for aggregating them into a single global DoB index.
This is an important limitation, since it is typically easier and more meaningful to compare
a scalar DoB value between different networks. Moreover, global measures, being scalar values,
are probably more useful for designing clustering or community detection methods.
}

Thus, several approximations have been proposed which can roughly be divided 
into two families of local and global measures. 
Local measures attain efficiency by focusing only on cycles of particular,
usually short, lengths, such as 3-cycles (triads). They can be fast, but provide only
a limited description of the real structure of a network. Thus, we argue that global measures 
are preferable.

Several global approaches have been proposed. Some bypass the problem
of counting cycles entirely, and instead search for partitions minimizing 
frustration~\cite{facchettiComputingGlobalStructural2011}
(the number or relative weight of edges incompatible with the SBT assumptions),
but they suffer from similar computational constraints due to their combinatorial nature.
Others leverage spectral properties of signed graphs and are therefore computationally efficient,
but measure only strong balance and quantify DoB using the smallest eigenvalue
of the signed Laplacian matrix~\cite{kunegisSpectralAnalysisSigned2010}, 
which is not normalized and can be difficult to compare between networks.
The last major approach is based on approximating cycle counts with counts of closed walks which can
be calculated, or at least approximated, very efficiently with standard linear
algebra~\cite{estradaWalkbasedMeasureBalance2014,singhMeasuringBalanceSigned2017}. 
Moreover, it can produce both local and global 
measures~\cite{estradaWalkbasedMeasureBalance2014,diaz-diazNetworkTheoryMeets2023} 
as well as capture strong and weak balance properties~\cite{kirkleyBalanceSignedNetworks2019}. 
However, walk-based approximations can be potentially
misleading as they may combine patterns found at very different cycle
lengths~\cite{giscardEvaluatingBalanceSocial2017}. 
\rev{
On the other hand, one can put forth arguments based on the theory of dynamical consensus 
on signed networks and argue that closed walks provide a fuller picture of structural 
balance~\cite{estradaRethinkingStructuralBalance2019}.
}

Here we propose Multiscale Semiwalk Balance (MSB): an approach applicable to
(un)directed, (un)weighted signed networks.
It is multiscale as it provides both grayscale measures approximating DoB at particular 
cycle lengths, as well as global indicators aggregating local measures across multiple scales 
in a principled manner. 
\rev{
Namely, it enforces what we call Locality Principle (LP) and ensures
that global DoB estimates are weighted averages of estimates at specific lengths such that
DoBs for shorter cycles are assigned with non-decreasing weights.
}

Our work builds on the Walk Balance (WB) approach proposed by 
Estrada and Benzi~\cite{estradaWalkbasedMeasureBalance2014},
\revII{which tends to underestimate DoB}, especially in large
networks~\cite{singhMeasuringBalanceSigned2017,giscardEvaluatingBalanceSocial2017}.
We show that this is caused by too much weight being placed on long cycles and can be fixed
by introducing a formal resolution parameter. Namely, we demonstrate how the inverse temperature, 
$\beta$, considered briefly already in Ref.~\cite{estradaWalkbasedMeasureBalance2014}, 
can be reinterpreted and used to determine 
\rev{
an appropriate weighting scheme for aggregating DoB measures across different cycle lengths
that satisfies LP.
}
It also allows our MSB approach to be applicable and meaningful in the context of weighted signed networks.
Additionally, we generalize the WB approach to capture both strong and weak balance,
as well as define DoB measures not only at the level of entire graphs but also for particular nodes 
and pairs of nodes to enable the development of effective SBT-aware clustering (community detection)
methods.
\rev{
Last but not least, by using semiwalk-based approximations our methods are more directly linked to
both undirected and directed SBT theorems and therefore meaningful also for directed signed networks.
}

We demonstrate the utility of our approach 
in two case studies of polarization in social systems. The first is a re-analysis
of the famous Sampson's Monks dataset~\cite{sampsonNovitiatePeriodChange1968},
in which we show that the commonly accepted \enquote{ground truth} partition is not SBT-optimal
by finding better ones, which also shed some additional light on the underlying social dynamics. 
In the second study we use our methods to provide evidence for increasing
polarization in the U.S. Congress based on bill co-sponsorship data~\cite{nealSignTimesWeak2020}.

\subsection{Notation}\label{sec:intro:notation}

Here we consider weighted graphs $G = (V, E, \omega)$ with $n = |V|$ vertices and $m = |E|$
edges and no self-loops or multilinks, where $V$ and $E \subseteq V \times V$ 
are vertex and edge sets respectively, 
and $\omega: E \to \mathbb{R}$ is a function assigning weights to edges. 
The weights can be negative, so the above definition captures all 
(un)signed, (un)weighted and (un)directed graphs.

\rev{
The adjacency matrix of a graph $G$ is given by a square $n \times n$ matrix $\mathbf{A}(G)$
such that $\mathbf{A}_{ij} = \omega_{ij} = \omega(i, j)$
if $(i, j) \in E$ or otherwise $\mathbf{A}_{ij} = 0$. 
Whenever possible without introducing ambiguity, we will drop the explicit dependence on $G$ 
and prefer a simpler notation, $\mathbf{A}$. We will use $|\mathbf{A}|$ to denote the unsigned
counterpart of $\mathbf{A}$ such that $|\mathbf{A}|_{ij} = |\omega_{ij}|$. 
Additionally, $\mathbf{P}$ and $\mathbf{N}$ will denote non-negative $n \times n$ matrices
corresponding to positive and negative parts of $\mathbf{A}$ 
such that $\mathbf{A} = \mathbf{P} - \mathbf{N}$ and $|\mathbf{A}| = \mathbf{P} + \mathbf{N}$.
When discussing network partitions we will use $\mathbf{B}$ to denote $n \times b$
block partition matrix such that $\mathbf{B}_{iu} = 1$ when the $i$th node belongs to the
$u$th block (group) or otherwise $\mathbf{B}_{iu} = 0$.
Matrix trace operator will be denoted by $\tr$. In particular, trace of the $k$th power
of a square matrix $\mathbf{X}$ will be denoted by $\tr\mathbf{X}^k$.
Hadamard (elementwise) matrix product will be denoted by $\odot$.

All measures that we will define here will depend on a particular graph $G$. 
Thus, for the sake of avoiding cluttering the notation, whenever possible,
we will omit this general dependence in the notation.
}

\section{Results}\label{sec:results}

\rev{
\subsection{Preliminaries}\label{sec:pre}

Before introducing the proposed framework we first state the core problems our work is supposed to solve
in a more formal fashion for the sake of clarity.
}

\rev{
\subsubsection{Aggregating DoB measures}\label{sec:results:pre:dob}

The difficulty with defining a meaningful global Degree of Balance (DoB)
can be easily seen by first considering DoB measures for cycles of particular lengths.
For a signed graph $G$ we define $k$-balance (DoB for cycles of length $k$) as:
\begin{equation}\label{eq:general-k-balance}
    B_k = \frac{\mu_+(k)}{\mu_+(k) + \mu_-(k)}
\end{equation}
where $\mu_+(k)$ and $\mu_-(k)$ are respectively counts of balanced and unbalanced
cycles of length $k$. This measure is easy to interpret, since it is concerned with only one 
specific class of cycles (those of length $k$), so in this context it is justified to treat 
every cycle equally.

However, defining a global DoB measure integrating structural balance information
across different cycle lengths is more difficult, since there are infinitely many ways to do it.
A reasonable solution is to assume that global DoB should be a weighted average of
$k$-balance scores:
\begin{equation}\label{eq:general-balance}
    B = \sum_{k}\omega_kB_k
\end{equation}
where $\omega_k$'s are normalized weights ($\omega_k \geq 0$ and $\sum_k\omega_k = 1$) 
assigned to different balance scores at different lengths $k$. 
However, it is not clear how the weights should be chosen in order to produce a meaningful 
global DoB measure.

Importantly, let us note that the above generic definitions are appropriate for both the strong 
and weak notions of balance. In what follows we will derive particular operationalizations of 
these generic formulas.
}

\rev{
\subsubsection{Finding clusters in signed networks}\label{sec:results:pre:findex}

While it is useful to know DoB of a network, which tells how close it is to being perfectly balanced
and therefore clusterable, it is arguably even more useful to be able to find clusters
(network communities) such that they agree with SBT to the greatest extent possible. 
This compatibility of a given partition of a signed
network with respect to the structure theorems of SBT 
(see Methods, Sec.~\ref{sec:methods:sbt}, for details)
can be measured with 
\revII{frustration ratio}, which can be defined, 
following Ref.~\cite{doreianPartitioningApproachStructural1996},
as the sum of absolute weights of negative in-group and positive out-group ties relative 
to the sum of all absolute edge weights, which can be expressed succinctly in the matrix form as:
\begin{equation}\label{eq:findex}
    F(\mathbf{B}) = \frac{%
        \mathbb{1}^\top\left[
        (\mathbf{BB}^\top) \odot \mathbf{N} + 
        (\mathbb{1}\mathbb{1}^\top-\mathbf{BB}^\top) \odot \mathbf{P}
        \right]\mathbb{1}
    }{%
        \mathbb{1}^\top|\mathbf{A}|\mathbb{1}
    }
\end{equation}
where $\mathbb{1}$ is a vector of ones of an appropriate length,
$\mathbf{B} \in \mathbb{R}^{n \times b}$ is a block-partition matrix
and $\mathbf{P}$ and $\mathbf{N}$ are positive and negative parts of the adjacency matrix $\mathbf{A}$.
\revII{
Note that frustration ratio can be also seen as a normalized version of frustration count,
which is used to define frustration index as the minimal frustration count over all partitions
of a network~\cite{arefBalanceFrustrationSigned2019}.
}

\revII{
Frustration ratio is a very straightforward measure of the extent to which a given
partition produces a balanced network configuration. 
It ranges from 0 for balanced partitions 
to 1 for maximally unbalanced ones (Fig.~\ref{fig:findex}).
}

\begin{figure}[ht!]
\centering
\includegraphics[width=\textwidth]{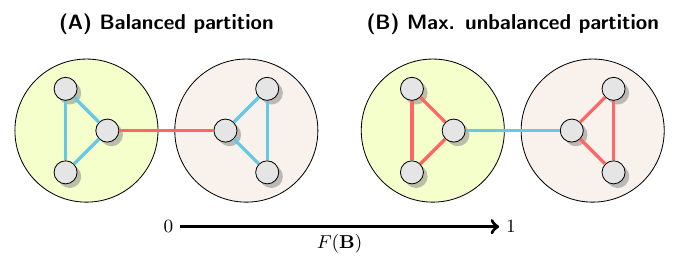}
\caption{
    Relationship between frustration ratio and structural balance in signed networks.
    Positive ties are blue and negative are red. Different groups are marked with circles.
}
\label{fig:findex}
\end{figure}
}

\rev{
It is important to note that frustration ratio, while closely related to DoB,
measures something different. DoB is a property of a network as such, which, thanks to the
structure theorems of SBT, is informative of the extent to which a given network is clusterable.
On the other hand, frustration ratio is a property of a network and a specific partition, and is directly
related to how close a given partition is to be perfectly balanced. That is why we argue that
it is an appropriate measure of the quality of a partition vis-à-vis the tenets of SBT.
Thus, DoB and frustration ratio are closely related but not
equivalent, as already observed in Ref.~\cite{estradaRethinkingStructuralBalance2019}.
However, the crux is that in the limiting case of the perfect balance, DoB equal to 1
implies that there is a partition with zero frustration and vice versa. 
The farther a network is from this ideal case the fuzzier this relationship gets, 
but in general the two measures will be always related.
We will use this insight to develop a clustering method utilizing DoB-like scores.
}

\rev{
\subsubsection{Approximating (semi)cycles with closed (semi)walks}\label{sec:results:pre:approx}

Counting cycles is computationally very expensive, so in practice approximations are necessary.
A very general and flexible approach is based on approximating cycles with closed walks, which can
be counted much more efficiently using powers of adjacency matrix. However, SBT in its most general
form applicable to both directed and undirected networks is formulated in terms of closed semipaths, 
or semicycles~\cite{cartwrightStructuralBalanceGeneralization1956}. 
A semipath is a path, in which edge directions can be ignored, 
but any edge can still be traversed only once.
This property has an important consequence for directed networks, in which in general semicycles
correspond to cycles in the associated undirected multigraph (obtained by making every link bidirectional)
with the exception of 2-cycles, which require both $i \to j$ and $j \to i$ links to be present
(Fig.~\ref{fig:semicycles}).

Thus, we argue that semicycle counts should be approximated using semiwalks, which are simply
walks on the corresponding undirected multigraph 
(i.e. ignoring edge directions)~\cite{wassermanSocialNetworkAnalysis1994}. 
However, an additional correction factor should be used to account for 
the fact that non-reciprocated directed edges do not generate any 2-semicycles.

\begin{figure}[ht!]
\centering
\includegraphics[width=\textwidth]{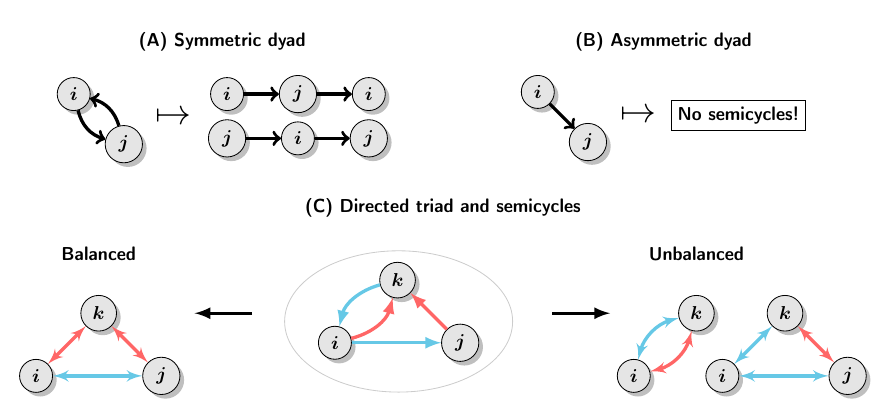}
\caption{
    Relationship between cycles and semicycles.
    \textbf{(A)}~Symmetric (reciprocated) dyads generate two (semi)cycles.
    \textbf{(B)}~Asymmetric dyads generate no (semi)cycles, since a semicycle of the form $i-j-i$
    would have to cross the directed $i \to j$ link twice.
    \textbf{(C)}~An example of the connection between directed cycles and semicycles
    in signed networks. A single directed triad can generate several different balanced and unbalanced
    2- and 3-semicycles (which here are marked with two-way arrows).
}
\label{fig:semicycles}
\end{figure}
}

\subsection{Multiscale Semiwalk Balance}\label{sec:results:swb}

\rev{
Here we introduce Multiscale Semiwalk Balance (MSB) approach which provides solutions to
all of the above-mentioned problems.
}
We first develop it without considering the role of edge weights, which, 
as we discuss later, appear in our approach naturally also in the context of unweighted networks. 
Once the core framework is established, we show that
it automatically extends to weighted graphs in a meaningful way.
Moreover, here we focus on the undirected version of MSB and strong balance.
In Sec.~\ref{sec:directed} we generalize our approach to directed signed graphs and
in Methods, Sec.~\ref{sec:methods:weak-balance}, to the weak notion of structural balance.

\rev{
In what follows we will use the fact that for a graph $G$ walks of length $k$ between nodes $i$
and $j$ are counted by the elements of the $k$-th power of its (unsigned) adjacency matrix, 
$|\mathbf{A}|^k$ 
(in the weighted case $|\mathbf{A}|^k$ gives weighted counts such that 
each walk is assigned a weight equal to the product over its constitutive edges). 
Importantly, such matrix powers can be calculated and approximated easily using eigendecomposition, 
especially for symmetric matrices and here we will use only such.

We will be particularly interested in weighted sums of matrix powers of the following form:
\begin{equation}\label{eq:W}
    \mathbf{W}(\mathbf{A}, \beta; k_{\min}, k_{\max})
    = \sum_{k=k_{\min}}^{k_{\max}}\frac{\beta^k}{k!}\mathbf{A}^k
    \approx e^{\beta\mathbf{A}}
\end{equation}
where $k$ iterates over a sequence of consecutive non-negative integers, $k_{\min}, \ldots, k_{\max}$,
and the second approximate equality is exact when $k_{\min} = 0$ and $k_{\max} = \infty$.
In what follows we will use a simpler notation, $\mathbf{W}(\mathbf{A}, \beta)$,
whenever it is clear from the context, or unimportant, what $k_{\min}$ and $k_{\max}$ are.
Moreover, any function depending on $\mathbf{W}(\ldots)$ is also implicitly parametrized
by $k_{\min}$ and $k_{\max}$ but we will omit this in the notation for the sake of brevity.
Note that here $\beta$ is a free parameter which can be used to control the weights assigned
to different powers of $\mathbf{A}$. We will use this fact later.
Moreover, both $\mathbf{W}(\mathbf{A}, \beta)$ and its trace can be approximated in an accurate
and efficient manner based on $m$ leading eigenvalues of $\mathbf{A}$
(see Methods, Sec.~\ref{sec:methods:perf}).
}

\subsubsection{Strong balance}\label{sec:results:strong-balance}

\rev{
Following Estrada and Benzi ~\cite{estradaWalkbasedMeasureBalance2014}, we note that 
powers of signed adjacency matrix, $\mathbf{A}^k,$ give differences between counts of 
positive and negative walks of a given length, while powers of unsigned adjacency matrix,
$|\mathbf{A}|^k$, count all walks of the given length. Thus, the sum of differences between 
weighted counts of positive and negative walks of a lengths $k = k_{\min}, \ldots, k_{\max}$
is given by $\mathbf{W}(\mathbf{A}, \beta)$. Similarly, $\mathbf{W}(|\mathbf{A}|, \beta)$ gives
the corresponding sum of weighted counts of all walks.

In the case of undirected networks considered here we have that $k_{\min} = 3$, 
since 2-cycles in undirected signed networks are always trivially balanced.
On the other hand, it should be that $k_{\max} \leq n$, since no cycle can be longer than
the number of nodes in a network, but it is not obvious what is the proper exact choice for $k_{\max}$.
However, any moderately large value will do, since the higher
order terms in Eq.~\eqref{eq:W} are quickly killed by the inverse factorial factor.
In Supplementary Information (SI), Sec.~\ref{app:sec:numerical}, we show that typically
$k_{\max} \geq 10$ is enough to get practically error-free results. However, to stay on the safe
side in all following analyses we always use $k_{\max} = 30$.
}

Counts of closed walks are given by the diagonal elements, so the overall counts are
given by appropriate matrix traces. Thus, to measure structural balance in a signed network one can
use Balance Index~\cite{estradaWalkbasedMeasureBalance2014}, 
or the ratio of the difference between weighted counts of balanced ($\mu_+$) and unbalanced ($\mu_-$)
closed walks to the weighted count of all closed walks: 
\rev{
\begin{equation}\label{eq:bindex-strong}
    R(\beta) 
    = \frac{\mu_+ - \mu_-}{\mu_+ + \mu_-}
    =\frac{\tr\mathbf{W}(\mathbf{A}, \beta)}{\tr\mathbf{W}(|\mathbf{A}|, \beta)}
\end{equation}
}

A conceptually simpler measure is Degree of Balance (DoB), proposed already 
by~\citet{cartwrightStructuralBalanceGeneralization1956}, which represents the proportion of balanced walks:
\rev{
\begin{equation}\label{eq:dob-strong}
    B(\beta)
    = \frac{\mu_+}{\mu_+ + \mu_-}
    = \frac{1}{2}\left[R(G, \beta) + 1\right]
\end{equation}
}

Following~\cite{estradaWalkbasedMeasureBalance2014} again, we can define node-level measures,
\rev{
also known as local balance~\cite{diaz-diazNetworkTheoryMeets2023},
}
simply by calculating diagonals instead of traces:
\rev{
\begin{align}
    r_i(\beta) 
    &= \frac{\mathbf{W}(\mathbf{A}, \beta)_{ii}}{\mathbf{W}(|\mathbf{A}|, \beta)_{ii}} 
    \label{eq:bindex-strong-node} \\
    b_i(\beta)
    &= \frac{1}{2}\left[r_i(\beta) + 1\right] \label{eq:dob-strong-node}
\end{align}
}
Note that we use lowercase letters to denote quantities describing individual nodes instead of the
global properties of entire graphs. We will follows this convention also when defining other node-level
measures.

Measures of $k$-balance (DoB at a particular length $k$) can also be easily defined:
\rev{
\begin{align}
    R_k &= \frac{\tr\mathbf{A}^k}{\tr|\mathbf{A}|^k} \label{eq:bindex-strong-k} \\
    B_k &= \frac{1}{2}(R_k + 1) \label{eq:dob-strong-k}
\end{align}
}
Note that these measures do not depend on $\beta$, since, even if they did, 
the same weighting factor would have to appear in both the numerator and denominator. 
This shows that $\beta$ indeed controls only the amount of weight put on different cycle
lengths, but does not influence the degree of balance at particular lengths.

\subsubsection{Contribution profiles and Locality Principle}\label{sec:results:contrib}

Importantly, one can asses the contribution of closed walks of length $k$ to the total 
weighted sum of closed walk counts for lengths $k_{\min}, \ldots, k_{\max}$:
\rev{
\begin{equation}\label{eq:contribution}
    C_k(\beta)
    = \frac{\beta^k}{k!} 
      \times \frac{\tr|\mathbf{A}|^k}{\tr\mathbf{W}(|\mathbf{A}|, \beta)}
\end{equation}
}

In other words, Eq.~\eqref{eq:contribution} measures the ratio of the weighted sum of closed walks
of length $k$ to the total weighted sum of closed walks over a specified range of lengths. 
It is normalized by construction, so $C_k(\beta) \in [0, 1]$ and 
$\sum_kC_k(\beta) = 1$.

The contribution score clearly depends on $\beta$, which can be used for 
controlling the influence of different length scales on the overall calculations. 
\rev{
This is a crucial feature of our approach as it allows for a straightforward operationalization
of Locality Principle (LP): shorter cycles should generally matter no less than longer ones.

\begin{definition}[Locality Principle]\label{def:lp}
    A graph $G$, a resolution parameter $\beta > 0$ and a sequence
    of consecutive integers $2 \leq k_{\min}, \ldots, k_{\max}$ satisfy
    Locality Principle if and only if the following set of inequalities holds:
    \begin{equation*}
        C_{k_{\min}}(\beta) \geq \ldots \geq C_{k_{\max}}(\beta) 
    \end{equation*}
\end{definition}

Thus, LP allows for identification of a range of \enquote{reasonable} values of $\beta$,
which is given by a set $(0, \beta_{\max}]$, where $\beta_{\max} > 0$ is the largest
value still satisfying LP. Crucially, $\beta_{\max}$ always exists for graphs that contain
at least one closed walk for lengths $k_{\min}, \ldots, k_{\max}$.

\begin{theorem}\label{thm:beta-max}
    Let $2 \leq k = k_{\min}, \ldots, k_{\max}$ be a sequence of consecutive integers
    and $G$ a graph such that $\tr|\mathbf{A}|^k > 0$ for all $k$'s.
    Then, there exists a value $\beta_{\max}$ such that Def.~\ref{def:lp} holds for values 
    $0 < \beta \leq \beta_{max}$ and does not hold for values $\beta > \beta_{\max}$.
\end{theorem}

\begin{proof}\label{proof:beta-max}
    Using Eq.~\eqref{eq:contribution} the condition for LP can be rewritten as: 
    \begin{equation*}
        \frac{\beta^k}{k!}\tr|\mathbf{A}|^k \geq \frac{\beta^{k+1}}{(k+1)!}\tr|\mathbf{A}|^{k+1}
    \end{equation*}
    which after some straightforward algebra gives the following condition for $\beta$:
    \begin{equation*}
        \beta \leq (k+1)\frac{\tr|\mathbf{A}|^k}{\tr|\mathbf{A}|^{k+1}}
    \end{equation*}
    Now we note that the right-hand side of the above inequality is always positive,
    so there is a maximal value $\beta_{\max} > 0$ satisfying all inequalities:
    \begin{equation*}
        \beta_{\max} \coloneqq \min_k{(k+1)\frac{\tr|\mathbf{A}|^k}{\tr|\mathbf{A}|^{k+1}}}
    \end{equation*}
    As a result, a $\beta$ value satisfies LP if and only if $\beta \in (0, \beta_{\max}]$,
    which ends the proof.
\end{proof}
}

\begin{figure}[htb!]
\centering
\includegraphics[width=\textwidth]{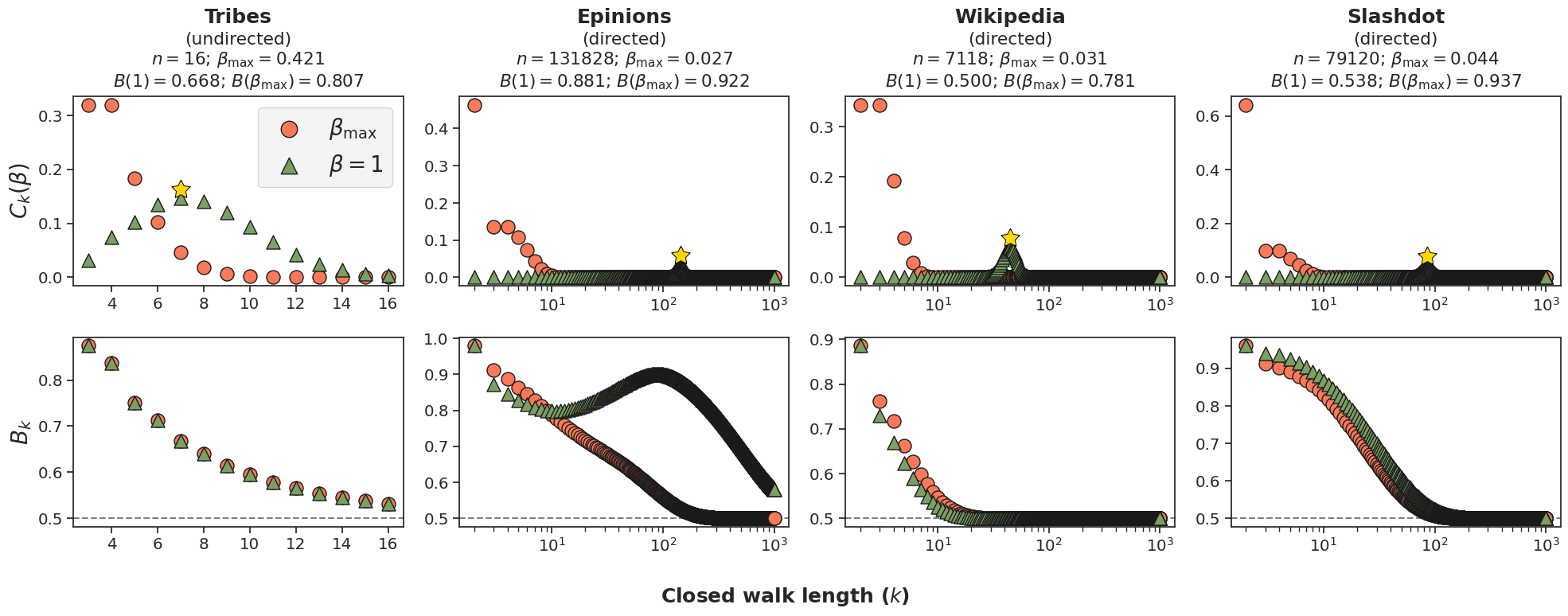}
\caption{%
    Contribution (top) and $k$-balance (bottom) profiles in four real networks studied by 
    Estrada and Benzi~\cite{estradaWalkbasedMeasureBalance2014}
    (see Methods, Sec.~\ref{sec:methods:datasets}, for dataset descriptions)
    based on WB ($\beta = 1$) and MSB ($\beta_{\max}$) approaches.
    Approximations based on $m = 10$ leading eigenvalues from both ends of the spectrum were used.
}
\label{fig:contrib}
\end{figure}

Finally, following the parsimony principle, we choose the weakest LP assumption possible and set
$\beta \coloneqq \beta_{\max}$. 
\rev{
This is a simple heuristic and we do not make any claims regarding its optimality.
We chose to use it here as developing a more principled method for selecting $\beta$
is beyond the scope of this paper and we plan to address this problem in the future.
However, as we later show through empirical analyses of real-world datasets, this heuristic
seems to work very well in practice. Moreover, using $\beta_{\max}$ still yields
markedly right-skewed contribution profiles, 
even though it can be argued that for this choice LP
\enquote{barely} holds, but this is true only in the sense of the entire set of inequalities 
for all pairs of lengths $(k, k+1)$, and does not imply that contribution scores assigned to short
closed walks are only marginally higher than those assigned to long walks (cf.~Fig.~\ref{fig:contrib}).
}

Our results also explain why the original WB approach~\cite{estradaWalkbasedMeasureBalance2014}
underestimates DoB in large networks. 
Namely, it does so because without determining the characteristic scale of a network by tuning $\beta$
the contribution profile may peak over very long cycles. 
\rev{
As Fig.~\ref{fig:contrib} shows,
WB places most of the weight on very long cycles ($k \approx 100$) in large networks, 
which clearly violates LP. As a result, it produces much lower DoB estimates than MSB, 
since products of signs over very long closed walks are arguably mostly random. 
Only in the case of the directed Epinions network WB produces an estimate close to the one given by MSB. 
However, as balance measures at particular cycle lengths show, this happens only because of the very
particular structure of the network resulting in high DoB at cycle  lengths of approximately 100.
Moreover, this seems to be a statistical artifact which disappears  almost completely 
when balance is assessed based on semiwalks (MSB) instead of ordinary walks (WB)
(see Sec.~\ref{sec:directed} for the generalization to directed measures based on semiwalks).
}
Crucially, this problem is likely to affect any other walk-based methods, which do not use a well-tuned
resolution parameter. Moreover, without a measure akin to Eq.~\eqref{eq:contribution}, 
it is hard to know for sure whether a given method will produce correct results for a given network.

\rev{
Importantly, global DoB is a weighted average of $k$-balance values with weights equal 
to the corresponding contribution scores. Thus, Eq.~\eqref{eq:dob-strong} satisfies the requirement
postulated in Sec.~\ref{sec:results:pre:dob}.

\begin{theorem}\label{thm:strong-dob-average}
    Let $G$ be a signed graph, $\beta > 0$ a resolution parameter and 
    $2 \leq k = k_{min}, \ldots, k_{\max}$ a sequence of consecutive integers. Then:
    \begin{equation*}
        B(\beta) = \sum_{k}C_k(\beta)B_k
    \end{equation*}
\end{theorem}

\begin{proof}\label{proof:strong-dob-average}
    It is given in the SI, Sec.~\ref{app:sec:strong-dob-average}.
\end{proof}
}

\rev{
\subsubsection{Node contributions}\label{sec:results:contrib-node}

Starting from similar ideas, one can also define node-level, or local, contribution scores
measuring the influence of a node $i$ on the overall DoB calculations:
\begin{equation}\label{eq:contrib-node}
    c_i(\beta) = \frac{\mathbf{W}(|\mathbf{A}|, \beta)_{ii}}{\tr\mathbf{W}(|\mathbf{A}|, \beta)}
\end{equation}
Note that by construction $c_i(\beta) \in [0, 1]$ and $\sum_ic_i(\beta) = 1$, so it enjoys
the same normalization property as the global contribution score. Importantly, node-level contribution
scores, together with local DoB, can be useful for defining and measuring various notions of node
centrality in signed networks.
}

\rev{
\subsubsection{Pairwise cohesion and clustering}\label{sec:results:pairwise}

Note that off-diagonal elements of $\mathbf{W}(\mathbf{A}, \beta)$ also convey important information.
Namely, they measure the difference between weighted counts of positive and negative walks between
nodes $i$ and $j$. We use this fact to define pairwise Cohesion Index:
\begin{equation}\label{eq:cindex}
    r_{ij}(\beta) = 
    \frac{\mathbf{W}(\mathbf{A}, \beta; k_{\min} = 2)_{ij}}%
    {\mathbf{W}(|\mathbf{A}|, \beta; k_{\min} = 2)_{ij}}
\end{equation}
and a corresponding (pairwise) Degree of Cohesion (DoC) measuring the fraction of positive walks
between nodes $i$ and $j$:
\begin{equation}\label{eq:doc}
    b_{ij}(\beta) = \frac{1}{2}\left(r_{ij}(\beta) + 1\right)
\end{equation}
Note that cohesion index uses $k_{\min} = 2$. This facilitates differentiating between
frustrated and non-frustrated edges. If there are many positive walks between $i$ and $j$,
but $i \sim j$ edge is negative, then the $(i, j)$ pair generates many unbalanced closed walks
and therefore the $i \sim j$ edge should be considered rather a frustrated in-group tie
than an out-group tie, and an analogous argument can be made for negative walks.
Thus, direct links by themselves do not provide evidence necessary for partitioning nodes
and therefore should not be used for determining pairwise cohesion.

We use the same letters $r$ and $b$ we used to denote (local) balance measures for the sake of consistency
as balance and cohesion are based on the same idea. Indeed, all balance scores 
can be seen as measures of \enquote{self-cohesion}. To see this, let us consider a cycle and a node
$i$ that sends a bit of information to its left neighbour, who passes it further to its left neighbour
and so on, until the bit comes back to $i$. Moreover, let us assume that the bit is flipped when
crossing negative edges. Now, it is easy to see that the bit will return in the original state
if and only if the cycle is balanced. In this sense, structural balance is measuring the consistency
between sent and returning signals.

Cohesion measures are important because they allow developing SBT-aware clustering methods.
We leave a detailed study of this idea for future work. However, in what follows we combine them
with standard agglomerative hierarchical clustering~\cite{hastieElementsStatisticalLearning2008}
(see Methods, Sec.~\ref{sec:methods:hclust}, for details)
to show that MSB approach produces meaningful results 
\revII{
and allows for detecting interpretable low frustration network partitions.
}
}

\subsubsection[Weighted measures and \textbeta{} as average edge weight]{%
    Weighted measures and $\beta$ as average edge weight
}\label{sec:results:weighted}

Importantly, $\beta$ can be interpreted in terms of an average edge weight. Any unweighted network can be seen as a weighted network with uniform absolute edge weights of 1. Note that in this case the absolute product over a closed walk of any length is always equal to 1, so every walk is considered equal, 
and it is only $\beta$ that controls and re-scales edge weights inducing nonuniform walk weights 
(through $\beta^k$ scaling). Thus, an arguably natural way to handle
non-unitary weights is to re-scale them, so the average absolute weight is equal to 1:
\begin{equation}\label{eq:weighting-scheme}
    \omega'_{ij} = \frac{|E|\omega_{ij}}{\sum_{kl} |\omega_{kl}|}
\end{equation}
where $\omega_{ij}$ is the original weight of the $(i, j)$ edge and $|E|$ is the number of edges.

This retains the interpretation of $\beta$ in terms of an average edge weight and ensures that in a
network with a completely \rev{uniformly random} topology 
(i.e.~Erdős–Rényi random graph with randomly and independently assigned signs and absolute weights)
the expected value of a walk weight (i.e. the product of the
corresponding edge weights) gets fixed to 1 when $\beta = 1$.
\revII{
Analyses in Sec.~\ref{sec:results:monks} suggest tentatively that this approach to incorporating 
edge weights may be indeed effective and produce better results than analogous unweighted methods
(e.g.~find partitions with lower frustration).
}

\revII{
\subsection{Directed measures}\label{sec:directed}

Here we extend all the previously defined measures to directed signed networks.
To do so, we first note that the structure theorems of SBT in their most general form
are formulated in terms of semipaths and semicycles (they are listed in Methods, Sec.~\ref{sec:methods:sbt}).
Thus, it is natural to extend our approach to directed networks by simply using semiwalks
instead of ordinary walks.

\begin{definition}[Semiwalk]\label{def:semiwalk}
    A semiwalk is a sequence of adjacent edges such that for every two consecutive edges 
    $(i, j)$ and $(k, l)$ it holds that $k \in \{i, j\}$ or $l \in \{i, j\}$.
\end{definition}

More intuitively, semiwalks are just ordinary walks ignoring edge
directions~\cite{wassermanSocialNetworkAnalysis1994}, or walks on an undirected multigraph derived
from a given directed graph by making all edges bidirectional. Thus, semiwalks between all pairs of nodes
in a graph $G$ are counted by powers of its semiadjacency matrix, which is defined as the symmetric part 
of the adjacency matrix:
\begin{equation}\label{eq:semiadjacency}
    \mathbf{S}(\mathbf{A}) = \frac{1}{2}\left(\mathbf{A}+ \mathbf{A}^\top\right)
\end{equation}
Note that $\mathbf{S}$ is symmetric and $\mathbf{S}(\mathbf{A}) = \mathbf{A}$ when $\mathbf{A}$ is symmetric,
which jointly means that $\mathbf{S}[\mathbf{S}(\mathbf{A})] = \mathbf{S}$, so the semiadjacency operator
is idempotent. In what follows, we will use a simpler notation without the explicit dependence on $\mathbf{A}$
and we will use $\mathbf{S}$ to denote $\mathbf{S}(\mathbf{A})$ and $|\mathbf{S}|$ to denote
$\mathbf{S}(|\mathbf{A}|)$.

\begin{figure}[ht!]
\centering
\includegraphics[width=\textwidth]{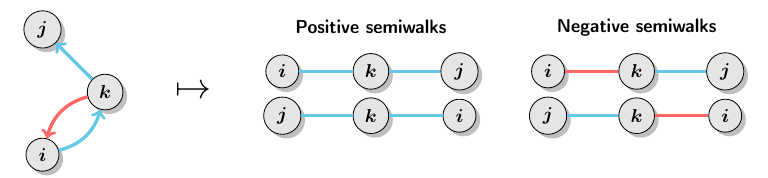}
\caption{
Semiwalks in directed signed networks. Positive and negative semiwalks passing through symmetric
dyads with opposite edge signs cancel each other out.
}
\label{fig:semiwalks}
\end{figure}

Importantly, $\mathbf{S}$ is not a lossless representation of the adjacency matrix
of the undirected multigraph underlying a given directed signed network, but it is lossy in a way 
which does not affect any balance-related calculations. 
Firstly, reciprocal edges with opposite signs cancel each other out in $\mathbf{S}(\mathbf{A})$. 
However, this does not affect the difference between counts of positive and negative semiwalks,
$\mu_+ - \mu_-$, since each symmetric dyad with opposite edge signs will be included in the same
number of positive and negative semiwalks between $i$ and $j$ (Fig.~\ref{fig:semiwalks}).
Secondly, the $1/2$ factor means that $\mathbf{S}$ approximates the adjacency matrix of the 
multigraph divided by 2,
but, again, this does not matter as in our approach edge weights are reweighted by the $\beta$ parameter,
which sets the average edge weight, anyway. The gain from using the $1/2$ factor is that $\mathbf{S}$ is
idempotent and equal to $\mathbf{A}$ for undirected graphs.

As a result, directed balance measures are obtained simply by substituting $\mathbf{A}$
with $\mathbf{S}$ and $|\mathbf{A}|$ with $|\mathbf{S}|$ in all the formulas. However, to account
for the fact that 2-cycles in directed signed networks are not trivial
(i.e.~they may be both balanced and unbalanced), an additional correction
is needed. As explained in Sec.~\ref{sec:results:pre:approx}, asymmetric dyads do not span any
2-semicycles, while symmetric ones do. Thus, in the case of directed networks one needs to apply
corrections to Eqs. \eqref{eq:W} and \eqref{eq:V} to count proper 2-semicycles:
\begin{align}
    \vec{\mathbf{W}}(\mathbf{A}, \beta) 
    &= \frac{\beta^2}{2}\mathbf{A}^2 + \mathbf{W}(\mathbf{S}, \beta)
    \label{eq:W-dir} \\
    \vec{\mathbf{V}}(\mathbf{A}, \beta)
    &= \frac{\beta^2}{2}\left(\mathbf{PN} + \mathbf{NP}\right) + \mathbf{V}(\mathbf{S}, \beta)
    \label{eq:V-dir}
\end{align}
where both $\mathbf{W}$ and $\mathbf{V}$ still use $k_{\min} = 3$.
}

\subsection{Re-analysis of Sampson's Monastery dataset}\label{sec:results:monks}

Sampson's Monastery study~\cite{sampsonNovitiatePeriodChange1968} produced one of the most famous network
datasets studied in Social Network Analysis (SNA) in general, and SBT in particular. It describes
the evolution of the social structure in a group of postulants and novices in a monastery 
in New England in 1960's. 
Namely, a network of liking (positive) and disliking (negative) relations was
measured at five points in time.
\rev{
The ties are directed and weighted in the $-3:3$ range, with weights indicating ordinal ranking of
the preference towards or against a given person typical for sociometric studies
(see Methods, Sec.~\ref{sec:methods:datasets:monks}, for details).
}
The dataset is particularly valuable because, as the study had been conducted, the group went through a major conflict, which eventually lead to either resignation or expulsion of the
majority of the members of the congregation. Moreover, Sampson identified a partition into three groups,
which later have been independently validated with analytic SBT-motivated
clustering methods~\cite{doreianPartitioningApproachStructural1996},
and therefore is commonly recognized as the \enquote{ground truth} solution.

\begin{figure}[ht!]
\centering
\includegraphics[width=\textwidth]{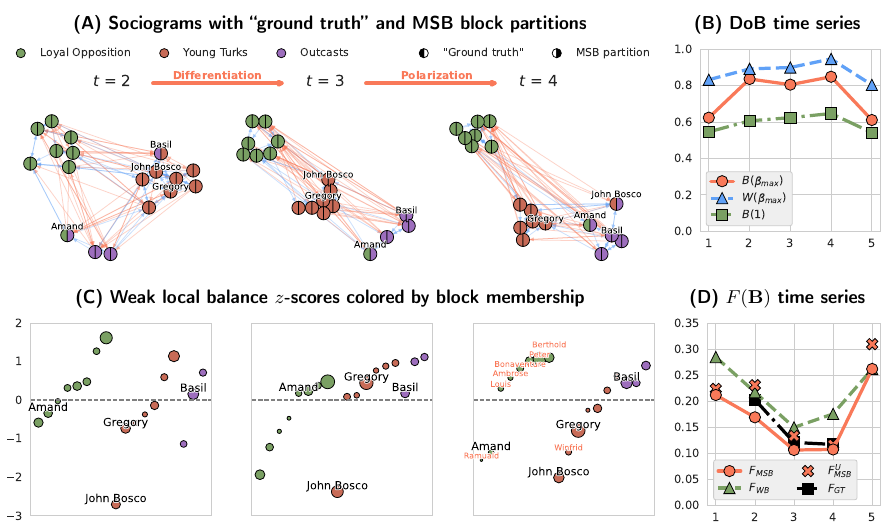}
\caption{
Re-analysis of Sampson's Monastery networks using MSB approach.
Full spectra were used in computations (exact results).
\textbf{(A)}~Signed sociograms at times $t=2, 3, 4$. Left side colors denote block membership according
to the \enquote{ground truth} partition and right side colors correspond to MSB partitions.
Positive ties are blue and negative are red. Individuals of which \enquote{ground truth} and MSB block
memberships differ (Amand, Basil and John Bosco) as well as the leaders of Young Turks 
(John Bosco and Gregory) are labelled. 
Network layout was determined with Kamada-Kawai algorithm using only positive ties with weights
(distances) on cross-block ties rescaled by the factor of 5.
\textbf{(B)}~Time series of strong and weak DoB measures for $t = 1, \ldots, 5$ 
using MSB as well as strong DoB
based on WB approach of \citet{estradaWalkbasedMeasureBalance2014}, which is equivalent to MSB approach
with $\beta=1$ using ordinary adjacency matrix.
\textbf{(C)}~Weak local balance expressed as $z$-scores relative to the overall distribution.
Points are sized proportionally to local contributions and ordered first by block membership and
then by balance scores. Members who remained at the monastery after the
culmination of the conflict ($t=5$) are marked with red labels on the subplot for $t=4$.
\textbf{(D)}~Time series of frustration ratios for $t = 1, \ldots, 5$
according to partitions obtained with MSB and WB ($\beta=1$) approaches as well as the 
\enquote{ground truth} solution (which is defined only for times $t=2,3,4$).
\rev{$F_{\text{MSB}}^U$ denotes frustration values using unweighted MSB approach.}
}
\label{fig:monks}
\end{figure}

The most important events happened at times $t=2,3,4$, which correspond to a period of differentiation
and polarization~\cite{doreianPartitioningApproachStructural1996} that eventually lead to an open
conflict and disintegration of the group. At $t=2$ twelve new members joined the monastery,
while some older members left after $t=1$, so the new group consisted of 18 men in total.
This perturbation lead to an emergence of two competing groups (Loyal Opposition and Young Turks)
as well as a group of peripheral members, who were not fully accepted by the rest (Outcasts).
The network at time $t=4$ depicts the structure just before the open conflict and disintegration.
At $t = 5$ only 7 members remained in the monastery, and those who stayed
(they are marked with red labels on Fig.~\ref{fig:monks}C, $t=4$)
belonged almost exclusively to the Loyal Opposition, which clearly \enquote{won} the conflict.

Here we use MSB approach to demonstrate that the \enquote{ground truth} partition 
is not SBT-optimal, or maximally consistent with Theorem~\ref{thm:weak-structure-I}.
This can be measured using frustration ratio, $F(\mathbf{B})$.
Fig.~\ref{fig:monks}A shows both the \enquote{ground truth} and the MSB network partitions 
for times $t=2,3,4$ (see Methods, Sec.~\ref{sec:methods:hclust}, for details of the clustering method).
They differ only in a few details, which are, nonetheless, very informative about the unfolding dynamics.
Firstly, according to the \enquote{ground truth} partition, Basil was a member of the Outcasts.
However, MSB analysis indicates that initially ($t=2$) he interacted mostly with the Young Turks
and only later was rejected and became one of the Outcasts. Secondly, Amand, a member of the 
Loyal Opposition according to the \enquote{ground truth}, was consistently identified as one of the
Outcasts by our MSB clustering procedure. Most importantly, according to MSB, John Bosco, 
who was considered one of the two leaders of the Young Turks (the second one was Gregory), became one of
the Outcasts just before the disintegration of the monastery ($t=4$). This says a lot about
why the Young Turks \enquote{lost} the competition against the Loyal Opposition, of which core constituted
most of the group that remained at the monastery. 

As evident in Fig.~\ref{fig:monks}C,
local weak balance scores of John Bosco were consistently low and at time $t=4$ also Gregory, the second
leader, attained low local balance
\rev{
(see Methods, Sec.~\ref{sec:methods:weak-balance} for the details of the weak balance measures).
}
This was largely driven by the tension in their personal relationship 
(at $t=4$ the Gregory$\to$John Bosco tie is positive and John Bosco$\to$Gregory is negative),
which then propagated through the entire group 
(note that both of them had high local contribution scores, Fig.~\ref{fig:monks}C)
leading, probably, to its decomposition. 
As Fig~\ref{fig:monks}A shows, over time John Bosco established more positive connections with Outcasts
and developed negative feelings towards Gregory. At the same time, the core of Loyal Opposition 
strengthened internal connections and became very cohesive at time $t=4$, as indicated by high weak 
local balance scores of most of the individuals with red labels on Fig.~\ref{fig:monks}.

Importantly, MSB measures of DoB are clearly high during the evolution of the
conflict ($t=2,3,4$), with the maximum at $t=4$, while analogous WP measures, 
which are not based on LP, yielded markedly lower DoB values
that cannot be readily interpreted as indicative of a conflict, as they are not much greater than
$1/2$ (which can be expected for a random assignment of edge signs). Similarly, frustration values
obtained with MSB clustering are consistently lower than those of \enquote{ground truth} partition,
and at times $t=1,2,3,4$ also lower than the ones obtained using WB.
\rev{
On the other hand, frustration ratios obtained when ignoring edge weights ($F_{\text{MSB}}^U$)
are markedly higher, indicating that our approach uses edge weights information effectively
leading to better results, i.e.~partitions with lower frustration.
}

\revII{
Thus, the analysis indicates that MSB can produce useful and interpretable results,
including finding low frustration partitions of signed networks.
}
Moreover, by combining global and local measures applied to time series of network snapshots,
insights into the impact of microscopic changes (e.g.~edge sign switching) on the meso- and macroscopic
structure can be gained. 

\subsection{Polarization in the U.S. Congress}\label{sec:results:congress}

It is often claimed that political life in contemporary democracies have polarized significantly
over the last few decades. Arguably, this debate is particularly relevant for the U.S.,
because of its largely two party political system, for which the notion of (bi)polarization is 
particularly well-defined. 
Such a hypothesis is also supported by a lot of empirical evidence
(cf.~\cite{nealSignTimesWeak2020,hohmannQuantifyingIdeologicalPolarization2023} and references therein).

\begin{figure}[thb!]
\centering
\includegraphics[width=\textwidth]{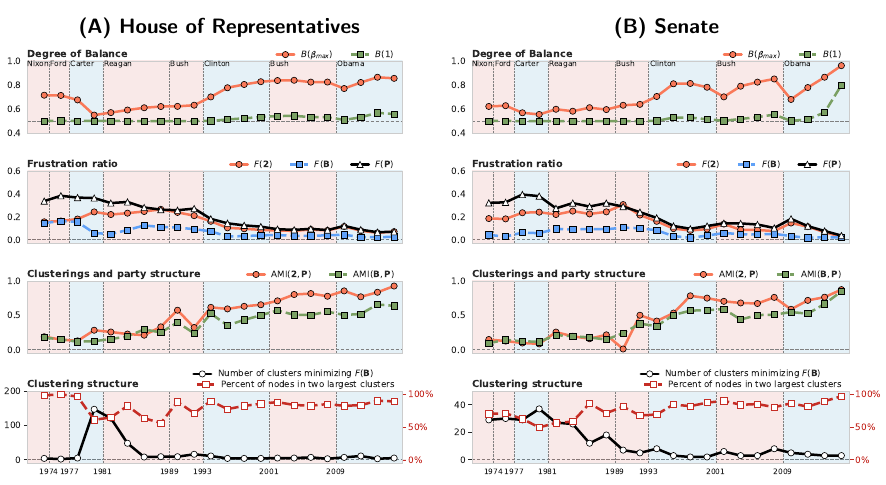}
\caption{
    Polarization in the U.S. Congress between 93th and 114th Congress (1973-2016).
    Panels are divided into regions corresponding to subsequent White House administrations
    with colors denoting Republican (red) and Democratic (blue) presidents.
    Approximations based on $m=10$ leading eigenpairs from the both ends of the spectrum were used.
    Starting from the top,
    \textbf{(1st~panel)} shows strong DoB time series based on MSB approach, $B(\beta_{\max})$ 
    and WB of \citet{estradaWalkbasedMeasureBalance2014}, $B(1)$.
    \textbf{(2nd~panel)} presents frustration ratios for best partitions into 2 clusters, 
    $F(\mathbf{2})$, general partitions minimizing $F(\mathbf{B})$,
    and partitions based on partisan affiliations, $F(\mathbf{P})$.
    \textbf{(3rd~panel)} quantifies similarity between party-based partitions and best bipartitions,
    $\text{AMI}(\mathbf{2}, \mathbf{P})$, as well as best partitions into $k$ clusters,
    $\text{AMI}(\mathbf{B}, \mathbf{P})$, using Adjusted Mutual Information
    (AMI) score~\cite{vinhInformationTheoreticMeasures2010}. The closer values are to 1, the better
    is the match between two clustering solutions.
    \textbf{(4th~panel)} shows the number of clusters in the solution minimizing $F(\mathbf{B})$
    (left $y$-axis), as well as the fraction of nodes within the two largest clusters (right $y$-axis).
}
\label{fig:congress}
\end{figure}

Here we use MSB approach to study polarization in both chambers of the U.S. Congress 
based on patterns of bill co-sponsorship between 1973 and 2016
(93rd to 114th Congress)~\cite{nealSignTimesWeak2020}.
The dataset consists of two sequences of \rev{undirected} signed networks
inferred from co-sponsorship data, where positive ties indicate statistically significant 
tendency of two representatives/senators to promote the same bills and negative ties the opposite
tendency to \revII{avoid promoting the same projects} 
(see~Methods, Sec.~\ref{sec:methods:datasets:congress} for details).

Our analysis indicates that polarization increased markedly in both the House of Representatives
(Fig.~\ref{fig:congress}A) and the Senate (Fig.~\ref{fig:congress}B). This is evident in the steadily
increasing strong DoB values meaning that co-sponsorship networks became easier to bipartition in time.
\rev{
The increasing trend seems to materialize during the second Congress of Carter's administration
and be stable, notwithstanding some transient perturbations.
}
Interestingly, and consistently with our previous analysis
of the importance of Locality Principle, WB approach yielded almost exclusively very low DoB values,
and thus would not capture the true trend. This is, of course, the consequence of the violation of LP.

In both chambers frustration ratios clearly converge (Fig.~\ref{fig:congress}, 2nd panels) 
meaning that 
\revII{
best bipartitions and clusterings (in $k$ groups) based on MSB approach
(Methods, Sec.~\ref{sec:methods:hclust}),
}
as well as partitions following partisan affiliations are becoming more and more consistent with the
SBT theorems and therefore also similar. This is evident in the time series of
the similarity between MSB and partisan partitions
\rev{
measured with Adjusted Mutual Information (AMI) 
score~\cite{vinhInformationTheoreticMeasures2010}
}
(Fig.~\ref{fig:congress}, 3rd panels).
Moreover, even in $k$-clusterings with $k$ large, most of the nodes tend to belong to the two 
largest clusters, indicating, again, an increasingly bipolar structure organized along the party lines.
\rev{
Note that even in the extreme case of the House of Representatives during the 96th congress 
(second congress of Carter's administration),
where we found 147 distinct \enquote{clusters}, 271 or around 61\% of the representatives belong to the
two largest blocks, meaning that the rest of the clusters correspond to the other 171 representatives,
for whom the average cluster size was about $1.18$. Thus, in this period many
members of the congress were effectively functioning in-between the two main blocks,
and from the perspective of the clustering procedure they were outliers forming many small clusters,
very often composed of only one node. This result is consistent with the fact that this was a period
of the lowest polarization, for which the partisan cleavage should not be very pronounced.
}

\rev{
To sum up, the results point to a strong consistency between global DoB measures and quality of
optimal partitions. Namely, the higher DoB the lower the frustration of optimal partitions found by
our clustering algorithm. Moreover, the fact that in time all empirical partitions become more
and more similar to the partisan affiliations and the majority of nodes always belong to the two largest
clusters jointly mean that the MSB partitions we obtained are meaningful and consistent with
the partisan polarization hypothesis. In other words, we indeed find that in time it becomes easier
to find low frustration network partitions that largely overlap with partisan affiliations. Thus,
the patterns of cooperation between the senators and representatives become more and more constrained
by their party membership.
}

\section{Discussion}\label{sec:discussion}

Polarization is often considered a salient, and perhaps worrying, feature of contemporary 
societies~\cite{schweighoferWeightedBalanceModel2020,nealSignTimesWeak2020,arefDetectingCoalitionsOptimally2020,hohmannQuantifyingIdeologicalPolarization2023}.
It can result in a sharp divergence of popular beliefs or attitudes (ideological polarization)
as well as in-group favouritism and out-group hostility 
(affective polarization)~\cite{hohmannQuantifyingIdeologicalPolarization2023}.
Crucially, the latter implies clustering of social networks into 2 or more groups
with primarily positive in-group and negative out-group ties. This structural aspect of polarization
is studied in Structural Balance Theory (SBT), which links it to properties
of semicycles in signed networks and provides strict criteria for measuring 
polarization~\cite{cartwrightStructuralBalanceGeneralization1956,davisClusteringStructuralBalance1967}.
 
Here we introduced Multiscale Semiwalk Balance (MSB) approach for measuring both strong and weak
degree of balance (DoB), which is applicable to any kind of (simple) signed networks, including directed
and weighted ones. MSB is computationally efficient by approximating semicycles with semiwalks,
which can be counted using standard linear algebra, and defines DoB measures not only for entire graphs
but also specific nodes and pairs of nodes, which in turn allows for implementing effective
signed community detection methods motivated by SBT.
\rev{
Crucially, MSB is multiscale in the three following senses:
\begin{enumerate}
\item It proposes a principled way of aggregating multiple $k$-balance scores for particular cycle
lengths to produce a single global DoB estimate motivated by Locality Principle (LP).
The resolution of analysis, or the weighting scheme for aggregating $k$-balance scores, 
is controlled by a single parameter, $\beta$, which can be tuned based on first principles to capture 
the characteristic scale of a network at which its DoB should be assessed.
This is a crucial feature of our framework, as even though many other approaches apply some decaying
weights to longer cycles, typically the decay rate is fixed or controlled by a free parameter with no
principled way of selecting an appropriate
value~\cite{estradaWalkbasedMeasureBalance2014,singhMeasuringBalanceSigned2017,giscardEvaluatingBalanceSocial2017,arefMeasuringPartialBalance2018,kirkleyBalanceSignedNetworks2019}.
\item It provides methods for measuring strong and weak DoB for entire graphs, 
closed walks of particular lengths, individual nodes and pairs of nodes.
\item Thanks to the pairwise measures it facilitates development of methods 
for finding mesoscopic structures in signed networks, i.e.~clusters or groups of nodes with primarily
positive in-group and negative out-group ties. 
\end{enumerate}
}

\rev{
Unlike many other approaches to
SBT~\cite{estradaWalkbasedMeasureBalance2014,singhMeasuringBalanceSigned2017,kirkleyBalanceSignedNetworks2019}, 
MSB is formulated explicitly in terms of semiwalks as an approximation to semipaths and semicycles.
This connects it more directly to the structure
theorems~\cite{cartwrightStructuralBalanceGeneralization1956,davisClusteringStructuralBalance1967}, 
and as a result facilitates meaningful analyses of directed networks.
Crucially, semiwalk-based $k$-balance scores tend to be similar to values produced by cycle-based
$k$-balance methods introduced in Ref.~\cite{giscardEvaluatingBalanceSocial2017} 
(see Methods, Sec.~\ref{sec:methods:perf:walks}, for details).
Thus, the fundamental approximation on which our approach is based seems to introduce little extra
noise relative to cycle-based measures. Similarly, the error introduced by using only leading
eigenvalues and eigenvectors is also typically very small 
(SI,~Sec.~\ref{app:sec:numerical}).
On the other hand, being based on (semi)walks that can be counted easily using standard linear algebra,
MSB computations can be remarkably fast (SI,~Sec.~\ref{app:sec:efficiency}).

Furthermore, there are also theoretical reasons for preferring walk-based over cycle-based DoB
measures. First, let us note that in a signed graph all cycles are balanced if and only if all closed
walks are balanced, so for measuring perfect structural balance walk- and cycle-based DoB measures
are equivalent. Furthermore, in opinion dynamics (diffusion) on a signed graph two groups can reach
different consensus states if and only if the graph is balanced, but the diffusion process depends not
only on purely cyclic structures, but also on acyclic ones, as well as \enquote{artificial cycles}
produced by backtracking walks~\cite{estradaRethinkingStructuralBalance2019}.
Thus, it can be argued that partial DoB measures defined in terms of (semi)walks paint a fuller
picture of structural balance, especially as far as the interplay between network structure
and diffusion dynamics is considered.

Thus, our perspective is different from other works on multilevel assessment of structural balance
such as Ref.~\cite{arefMultilevelStructuralEvaluation2020},
which are focused exclusively on strong balance,
and in which microlevel DoB analysis is equated with the triad-level DoB, 
mesolevel with the cohesiveness of the network partitions as such
(which is fully compatible with our framework),
and finally macrolevel is equated with the line index (or frustration index), 
but computed only for partitions into two groups. 
\revII{
Furthermore, our approach tries to follow the structure theorems of SBT as closely as possible
given its approximate walk-based nature.
Directed MSB measures are based on semiwalks, and thus they ignore edge directions, 
except for the special case of dyads (2-cycles), in which directions of both edges are considered
(this is accounted for by corrections defined in Eqs. \eqref{eq:W-dir} and \eqref{eq:V-dir}).
This design choice follows directly from the
fact that SBT was formulated in terms of semicycles, which are simply cycles in which edge directions
are ignored as long as each edge is traversed at most once.
On the other hand, 
}
methods from Ref.~\cite{arefMultilevelStructuralEvaluation2020}
use edge direction information in a more complex fashion, which, of course, may be insightful but
is not necessary from the vantage point of SBT theorems and the problem of network partitioning.
}

\rev{
Locality Principle is justified not only by its usefulness as a heuristic guiding DoB methods,
but also by a long history of social and psychological
research. In particular, experimental research on perception of structural balance in social networks
indicates that people pay more attention to small scale
structures~\cite{zajoncStructuralBalanceReciprocity1965}.
This is in line with other seminal results stressing the importance of proximity 
(both physical and social) for social phenomena such as social impact 
theory~\cite{latane1981psychology} and Dunbar's numbers~\cite{hillSocialNetworkSize2003},
which are closely related to the fact that social networks tend to be sparse and composed
of ties that are localized within some physical and/or social
space~\cite{talagaHomophilyProcessGenerating2020}.
Moreover, studies of structural balance using alternative cycle-based methods show that real-world networks
tend to have a cycle length threshold after which $k$-balance scores suddenly decrease to random-like
values (around 0.5)~\cite{giscardEvaluatingBalanceSocial2017}. 
In other words, structural balance typically manifests itself at the level of small- and medium-sized
structures, so DoB measures should account for that. This is exactly what LP does.
}

\rev{
Importantly, $\beta$ can be endowed with a physical interpretation, which helps to explain its role 
as a resolution parameter.
Note that cohesion index defined in Eq.~\eqref{eq:cindex}, from which all other MSB measures may
be derived, can be approximated by a ratio of elements of two matrix exponentials, 
$r_{ij}(\beta) \approx (e^{\beta\mathbf{A}})_{ij} / (e^{\beta|\mathbf{A}|})_{ij}$,
and the exponential of a rescaled adjacency matrix, such as $\beta\mathbf{A}$, is known as communicability,
which is a general measure of connectedness defined in terms of the
weighted sums of walks of different lengths between pairs of
nodes~\cite{estradaCommunicabilityComplexNetworks2008}.
In this context, $\beta$ can be interpreted as the inverse temperature of a thermal bath in which
a network is submerged. More generally, the thermal bath may represent an \enquote{external situation},
e.g.~the level of agitation of the system, which manifests itself by rescaling edge weights
with the $\beta$ factor. As a result, when $\beta \to 0$ (hot regime), there is no communicability
between nodes, and when $\beta \to \infty$ (cold regime), then there is infinite communicability
between all pairs of nodes~\cite{estradaPhysicsCommunicabilityComplex2012}.
Note that in both cases the actual network topology ceases to matter.
Thus, network structure is accounted for in DoB calculations only for appropriately chosen
intermediate values of $\beta$, and in this context LP provides an effective heuristic for
fine-tuning $\beta$ and finding the most relevant range of cycle lengths at which DoB
should be assessed.
}

This stresses the importance of multiscale approaches to SBT and network science more generally. 
By linking structural balance to
communicability~\cite{estradaCommunicabilityComplexNetworks2008,estradaPhysicsCommunicabilityComplex2012},
our results suggest that, perhaps, other network descriptors defined
in terms of walks, or powers of adjacency matrices, 
such as multiscale network entanglement~\cite{ghavasieh2021unraveling},
can be informed by Locality Principle. 
Note that contribution scores defined in Eq.~\eqref{eq:contribution}, and used for operationalizing LP,
can be calculated for any, also unsigned, network. Thus, LP is a heuristic for determining
the characteristic intensity and length of internode correlations, and this determines the appropriate
weighting scheme for aggregating walk-based measures across multiple scales. 
More generally, our results contribute also to the research on the importance of local structures in
networks~\cite{miloNetworkMotifsSimple2002,mattssonFunctionalStructureProduction2021,talagaStructuralMeasuresSimilarity2022}.

\rev{
Our work, of course, does not come without limitations. Firstly, even though cohesion measures
defined in Eqs. \eqref{eq:cindex} and \eqref{eq:doc} seem to open up new possibilities for designing
clustering or community detection methods for signed networks, the actual clustering algorithm we used
here is rather naive. Developing more mature methods derived from first principles will not be an easy
task and we leave it for future work. Moreover, it can be argued that an even better approach for tuning
$\beta$ could be based on setting it to a value that minimizes frustration of the best partition.
However, a proper solution to this problem would require a solid theory-driven clustering method
parametrized by $\beta$, which we do not currently have, so the choice $\beta \coloneqq \beta_{\max}$
should be considered the best working heuristic for selecting an optimal value for $\beta$ for now,
but it should be replaced with more mature solutions once they arrive. 
Furthermore, even though some in-depth insights
regarding similarities and differences between cycle- and walk-based DoB measures vis-à-vis the
tenets of SBT have been offered by Estrada~\cite{estradaRethinkingStructuralBalance2019},
one can argue that the debate on whether the former or the latter should be preferred is not yet settled.
Perhaps, an interesting \enquote{middle ground} perspective could be gained by studying DoB measures
based on non-backtracking (Hashimoto) matrices~\cite{torresNonbacktrackingCyclesLength2019}?
}

\section{Methods}\label{sec:methods}

\rev{
\subsection{Overview of Structural Balance Theory}\label{sec:methods:sbt}

Here we state the main definitions and theorems of SBT concerned with bi-clusterability as formulated 
by Cartwright and Harrary~\cite{cartwrightStructuralBalanceGeneralization1956}.
We use the general formulation based on semipaths and semicycles, so the theorems are
applicable to both undirected and directed graphs. Thus, we first define semipaths and semicycles.

\begin{definition}[Semipath]\label{def:semipath}
    A semipath is a walk in which each (directed) edge can be traversed both ways
    but only once and each node is visited exactly once. 
\end{definition}

\begin{definition}[Semicycle]\label{def:semicycle}
    A semipath starting and ending at the same node
    (which in this case is allowed to appear twice).
\end{definition}

\begin{corollary}
    Notions of paths/cycles and semipaths/semicycles are equivalent in undirected graphs,
    since an undirected edge is treated in this context as two directed edges
    pointing in opposite directions.
\end{corollary}

\begin{definition}[Strong balance property]\label{def:strong-balance}
    A signed graph is balanced if and only if every semicycle it contains is positive
    (the product over all edge signs is positive).
\end{definition}

\begin{theorem}[Strong structure theorem]\label{thm:strong-structure-I}
    A signed graph is balanced if and only if its vertices can be partitioned
    into two subsets such that positive edges connect vertices from the same subset 
    and negative ones link vertices from different subsets.
\end{theorem}

The above results were later generalized  by~\citet{davisClusteringStructuralBalance1967}, 
who provided necessary and sufficient conditions for $b$-clusterability 
(where $b \geq 2$ is an unknown integer).

\begin{definition}[Weak balance property]\label{def:weak-balance}
    A signed graph is weakly balanced if and only if no semicycle contains exactly one negative edge.
\end{definition}

\begin{theorem}[Weak structure theorem]\label{thm:weak-structure-I}
    A signed graph is weakly balanced if and only if its vertices can be partitioned into $b$
    subsets such that positive edges connect vertices from the same subset 
    and negative ones link vertices from different subsets.
\end{theorem}
}

\rev{
\subsection{Weak balance}\label{sec:methods:weak-balance}

Following Ref.~\cite{kirkleyBalanceSignedNetworks2019} we define non-negative matrices 
$\mathbf{P}(\mathbf{A})$ and $\mathbf{N}(\mathbf{A})$ corresponding to positive and negative parts 
of signed adjacency matrix such that 
$\mathbf{A} = \mathbf{P} - \mathbf{N}$ and $|\mathbf{A}| = \mathbf{P} + \mathbf{N}$.
In what follows we will use the simpler notation without the explicit dependence on $\mathbf{A}$,
but it is important to remember that $\mathbf{P}$ and $\mathbf{N}$ are functions of $\mathbf{A}$.

Weak balance is defined in terms of the extent to which a network is free of cycles with exactly
one negative edge. This single negative link can be placed anywhere along a path starting at node $i$.
Hence, we first define a matrix counting weakly unbalanced walks of length $k$ between nodes $i$ and $j$
in a signed graph $G$ as:
\begin{equation}\label{eq:Vk}
\begin{split}
    \mathbf{V}_k(\mathbf{A}) 
    &= \sum_{l=1}^k\mathbf{P}^{l-1}\mathbf{NP}^{k-l} \\
    &= \sum_{l=1}^k\mathbf{Q\Lambda}^{l-1}\mathbf{Q}^\top\mathbf{NQ\Lambda}^{k-l}\mathbf{Q}^\top \\
    &= \mathbf{Q}\left[    
        \left(\sum_{l=1}^k\mathbf{L}(k,l)\right) \odot \mathbf{M}
    \right]\mathbf{Q}^\top
\end{split}
\end{equation}
where $\mathbf{Q\Lambda{}Q}^\top$ is the eigendecomposition of $\mathbf{P}$,
\revII{
$\mathbf{M}$ is a shorthand for the product $\mathbf{Q}^\top\mathbf{NQ}$
that appears in the middle of the second line,
} 
and $\mathbf{L}(k,l)_{ij} = \lambda_i^{l-1}\lambda_j^{k-l}$. Moreover, we used the fact that
$\mathbf{\Lambda}^{l-1}\mathbf{M\Lambda}^{k-l} = \mathbf{L}(k,l) \odot \mathbf{M}$.

Now, a matrix with weighted sums of counts of walks
of lengths $k = k_{\min}, \ldots, k_{\max}$ joining nodes $i$ and $j$ is given by:
\begin{equation}\label{eq:V}
\begin{split}
    \mathbf{V}(\mathbf{A}, \beta) 
    &= \sum_k\frac{\beta^k}{k!}\mathbf{V}_k(\mathbf{A}) \\
    &= \mathbf{Q}\left\{\left[
        \sum_k\frac{\beta^k}{k!}\sum_{l=1}^k\mathbf{L}(k,l)
    \right] \odot \mathbf{M}\right\}\mathbf{Q}^\top
\end{split}
\end{equation}

Next, we can use Eq.~\eqref{eq:V} to calculate the overall weighted sums of counts of unbalanced 
closed walks from appropriate traces:
\begin{align}
    \tr\mathbf{V}(\mathbf{A}, \beta)
    &= \sum_k\frac{\beta^k}{k!}\tr\mathbf{V}_k(\mathbf{A})
    \label{eq:weakly-unbalanced-walks} \\
    \tr\mathbf{V}_k(\mathbf{A})
    &= k\sum_{i=1}^m\lambda_i^{k-1}\mathbf{M}_{ii}
    \label{eq:weakly-unbalanced-walks-k}
\end{align}
where we used the fact that trace is invariant under cyclic permutations and $\mathbf{Q}$ is orthonormal.
The weighted sum of counts of closed walks at a node $i$ is similarly given by the diagonal elements,
$\mathbf{V}(\mathbf{A}, \beta)_{ii}$.

Now, Eqs.~\eqref{eq:W} and \eqref{eq:weakly-unbalanced-walks}
can be used to define the measure of the overall weak balance:
\begin{equation}\label{eq:dob-weak}
    W(\beta) 
    = 1 - \frac{\mu_W}{\mu_+ + \mu_-}
    = 1 - \frac{\tr\mathbf{V}(\mathbf{A}, \beta)}{\tr\mathbf{W}(|\mathbf{A}|,\beta)}
\end{equation}
where $\mu_W$ is the sum of weighted counts of weakly unbalanced closed walks.
Weak pairwise cohesion scores are given by ratios of individual matrix elements:
\begin{equation}\label{eq:weak-cohesion}
    w_{ij}(\beta) 
    = 1 - \frac{\mathbf{V}(\mathbf{A}, \beta; k_{\min} = 2)_{ij}}%
    {\mathbf{W}(|\mathbf{A}|,\beta, k_{\min} = 2)_{ij}}
\end{equation}
with local (node-level) weak DoB given by the diagonal elements, $w_{ii}(\beta; k_{\min} = 3)$.
Similarly, weak $k$-balance is given by considering only closed walks of a particular length $k$:
\begin{equation}\label{eq:weak-dob-k}
    W_k = 1 - \frac{\tr\mathbf{V}_k(\mathbf{A})}{\tr|\mathbf{A}|^k}
\end{equation}
Importantly, as in the case of strong balance, global weak DoB can be expressed as a weighted
average of weak $k$-balance with weights given by the corresponding contribution scores
(see SI, Sec.~\ref{app:sec:weak-dob-average}, for the proof).

Last but not least, the trace of the matrix series defined in Eq. \eqref{eq:V} used for counting 
unbalanced closed walks always converges, so it is well-defined. Note that:
\begin{equation}\label{eq:weakly-unbalanced-walks-convergence}
    0 \leq
    \sum_k\frac{\beta^k}{k!}\sum_{l=1}^k
    \tr\mathbf{P}^{l-1}\mathbf{N}\mathbf{P}^{k-l}
    \leq
    \sum_{k=0}^\infty\frac{\beta^k}{k!}\tr\left(\mathbf{P} + \mathbf{N}\right)^k
    = \tr{}e^{\beta|\mathbf{A}|}
\end{equation}
where it is known that the rightmost matrix exponential and its trace always converge, 
so the middle part of the inequality must converge too.
}

\subsection{Hierarchical clustering with pairwise DoB measures}\label{sec:methods:hclust}

Here we will use the following naive, yet effective, clustering procedure for signed networks
based on pairwise cohesion measures (see Secs. \ref{sec:results:pairwise} and \ref{sec:methods:weak-balance}).
Let $\mathbf{D}^S_{ij} = 1 - b_{ij}(\beta_{\max})$ and $\mathbf{D}^W_{ij} = 1- w_{ij}(\beta_{\max})$
be pairwise dissimilarity matrices
\rev{
(so $\mathbf{D}^S_{ii} = \mathbf{D}^W_{ii} \coloneqq 0$)
}
based on the notions of strong and weak balance respectively, 
and let $N_b$ be the maximum number of clusters one is willing to consider. Then, for $b = 1, \ldots, N_b$:
\begin{enumerate}
    \item Run Hierarchical Clustering (HC)~\cite{hastieElementsStatisticalLearning2008} algorithm
    for $b$ clusters using $\mathbf{D}^S$ as input and calculate frustration index according to 
    Eq.~\eqref{eq:findex} for the obtained block-partition matrix $\mathbf{B}$.
    \item Run HC for $b$ clusters using $\mathbf{D}^W$ as input and calculate the corresponding
    frustration index.
    \item Store the lower of the two frustration indices and its corresponding block partition.
\end{enumerate}
Finally, choose the partition with the lowest frustration index.

\rev{
\subsection{Accuracy of semiwalk-based approximations}\label{sec:methods:perf:walks}

\begin{figure}[ht!]
\centering
\includegraphics[width=\textwidth]{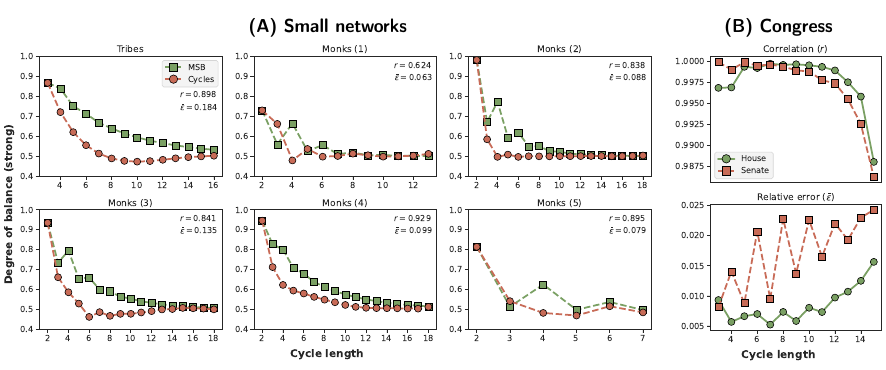}
\caption{
    Accuracy of semiwalk-based approximations relative to cycle-based DoB 
    estimates~\cite{giscardEvaluatingBalanceSocial2017} as measured by Pearson correlation
    $r$ and relative error $\bar{\epsilon} = \mean{|\frac{x - y}{y}|}$
    \textbf{(A)}~\enquote{Grayscale} ($k$-balance) measures based on semiwalks (MSB) and proper cycles
    in small networks studied in this paper
    (see Secs. \ref{sec:methods:datasets:tribes} and \ref{sec:methods:datasets:monks}). 
    In this case DoB values are reported for all possible cycle lengts and both MSB and cycle-based
    estimates are exact.
    \textbf{(B)}~Pearson correlations and relative errors for different cycle lengths calculated
    over co-sponsorship networks from the U.S. congress (Sec.~\ref{sec:methods:datasets:congress}).
    In this case MSB approximations are based on $m = 10$ leading eigenvalues and cycle-based
    estimates are approximated using sampling based on 10000 samples. Only cycles of lenght
    up to 15 were considered.
}
\label{fig:semiwalks-accuracy}
\end{figure}

MSB approach approximates semicycles with closed semiwalks. 
This is a fundamental design decision ensuring high computational efficiency,
but it comes at the price of introducing a discrepancy relative to cycle-based methods.
Here we present a comparison of $k$-balance methods provided by MSB and cycle-based approach from
Ref.~\cite{giscardEvaluatingBalanceSocial2017} based on several small and mid-sized networks.
The results indicate a strong similarity between the walk-based and the cycle-based DoB estimates
(Fig.~\ref{fig:semiwalks-accuracy}). 
Thus, it seems that the error introduced by walk-based approximations relative to cycle-based estimates
is typically small.
This should not come as a surprise as, thanks to Locality Principle, our MSB approach
ensures that DoB measures are driven primarily by patterns found in short closed walks, 
which coincide with cycles much more often than long walks
(e.g.~closed walks of length 3 are equivalent to 3-cycles).
}

\subsection{Numerical approximations and efficiency}\label{sec:methods:perf}

All computations of MSB can be implemented in a computationally efficient and accurate manner
using approximations based on $m$ leading eigenvalues and eigenvectors from both ends of the spectrum. 
Leading eigenpairs can be found very efficiently using modern linear algebra routines such as implicitly
restarted Arnoldi method~\cite{lehoucqARPACKUsersGuide1998,sorensenDeflationTechniquesImplicitly1996}.
Moreover, numerical stability can be guaranteed by conducting all computations in the log-space 
and using log-sum-exp trick (to avoid overflow when counting closed walks). 
This requires a bit of extra care as some eigenvalues may be non-positive. 
However, zero eigenvalues can be ignored altogether,
\rev{
since no measure defined here depends on the zeroth powers of adjacency matrices,
}
so the calculations can be done over the field of complex numbers, 
where the logarithm of any number with non-zero modulus is well-defined,
and cast back to real values only at the very end. As a result, MSB methods can be remarkably
efficient, even when applied to very large systems. 
\rev{
Secs. \ref{app:sec:numerical} and \ref{app:sec:efficiency} in the SI presents
empirical analyses of accuracy and efficiency of our implementation. Sec.~\ref{app:sec:analytic}
discusses the theoretical basis for approximations based on leading eigenvalues and eigenvectors. 
}

A more in-depth discussion of implementation details is beyond the scope of this paper, but we invite
the interested reader to study our source code (see: Data and code availability).

\subsection{Network datasets}\label{sec:methods:datasets}

\subsubsection{New Guinea Highlands tribes}\label{sec:methods:datasets:tribes}

An undirected unweighted signed network of friendships among tribes of Gahuku-Gama alliance structure 
of the Eastern Central Highlands region in New Guinea~\cite{readCulturesCentralHighlands1954}.
Edge sign indicates either friendship or enmity. 
Accessed from: \url{https://networks.skewed.de/net/new_guinea_tribes}

\subsubsection{Epinions trust network}\label{sec:methods:datasets:epinions}

This is a who-trust-whom online social network (directed, unweighted and signed) 
of a a general consumer review site \texttt{Epinions.com}. 
Members of the site can decide whether to \enquote{trust} each other. All the trust relationships interact
and form the Web of Trust which is then combined with review ratings to determine which reviews are shown
to the user~\cite{richardsonTrustManagementSemantic2003}.
Accessed from: \url{https://snap.stanford.edu/data/soc-Epinions1.html}.

\subsubsection{Wikipedia adminship vote}\label{sec:methods:datasets:wikipedia}

A directed unweighted signed network of votes on Request for Adminship (RfA) elections from a 2008 
snapshot of Wikipedia~\cite{leskovecGovernanceSocialMedia2010}.
Nodes represent editors, and a directed edge $(i,j)$ indicates that editor $i$ voted on editor $j$. 
Edge sign indicates the direction of the vote: positive = for, and negative = against. Edges are timestamped.
Accessed from: \url{https://networks.skewed.de/net/elec}.

\subsubsection{Slashdot Zoo network}\label{sec:methods:datasets:slashdot}

A directed unweighted signed network of interactions among users on Slashdot (\texttt{slashdot.org}), 
a technology news website~\cite{kunegisSlashdotZooMining2009}. 
Users name each other as friends (positive tie) or foe (negative tie). 
The friend label increases the scores of post, and the foe label decreases the score.
Accessed from: \url{https://networks.skewed.de/net/slashdot\_zoo}.

\subsubsection{Sampson's Monastery dataset}\label{sec:methods:datasets:monks}

Time series of 5 signed directed weighted networks measuring positive and negative relations between
postulants and novices in a New England monastery in 1960's~\cite{sampsonNovitiatePeriodChange1968}.
We used a version of the dataset studied in Ref.~\cite{doreianPartitioningApproachStructural1996}
in which edges have weights between -3 and 3 corresponding to the ranking of the least and most
(dis)liked/(dis)esteemed colleagues.
Accessed from: \url{http://vlado.fmf.uni-lj.si/pub/networks/data/esna/sampson.htm}.

\subsubsection{Co-sponsorship relations in the U.S. Congress}\label{sec:methods:datasets:congress}

Series of \rev{undirected} unweighted signed networks inferred from the data on bill co-sponsorships 
in both chambers of the U.S. Congress (House of Representatives and Senate) 
using Stochastic Degree Sequence Model~\cite{nealBackboneBipartiteProjections2014,nealSignTimesWeak2020}.
The data covers the period from 1973 (93rd Congress) to 2016 (114th Congress). Edges are signed, 
indicating the presence of a significant tendency to co-sponsor, or tendency to not co-sponsor, bills.
See SI, Sec.~\ref{app:sec:congress}, for a table with descriptive statistics.
Accessed from: \url{https://figshare.com/articles/dataset/A_Sign_of_the_Times/8096429}.
\section*{Data and code availability}

Sources of the data used in the paper are described in Methods.
The code and instructions for replicating the analyses, including a packaged Python code implementing
all MSB methods in a user-friendly manner, is available at Github (\url{https://github.com/sztal/msb}).
The repository provides also the datasets used in the analyses.

\section*{Acknowledgments}

S.T. acknowledges the support of National Science Center, Poland,
under a grant number \\ 2020/37/N/HS6/00796 (\textit{Outline of a network-geometric theory of social structure}). A.S.T. acknowledges support from FCT and the LASIGE Research Unit, ref.\ UIDB/00408/2020 and ref.\ UIDP/00408/2020.

\section*{Author contributions}

S.T. conceptualized the project, developed the mathematical framework
and its software implementation and conducted the analyses. S.T. and M.S. developed the physical
interpretation of the $\beta$ parameter. S.T, A.S.T. and T.J.S. worked out the theoretical
justification for the Locality Principle. S.T., M.S., T.J.S. and A.S.T. wrote the manuscript. 
S.T. reviewed and corrected the manuscript.

\section*{Competing interests}

The authors declare no competing interests.

\bibliographystyle{ACM-Reference-Format}
\bibliography{semiwalk-balance,shared-bibliography}

\newpage
\titleformat{\section}%
    {\centering\normalfont\Large\bfseries}{}{1em}{}
\renewcommand{\setthesubsection}{S\arabic{subsection}}
\setcounter{table}{0}
\renewcommand{\thetable}{S\arabic{table}}
\setcounter{figure}{0}
\renewcommand{\thefigure}{S\arabic{figure}}
\setcounter{equation}{0}
\renewcommand{\theequation}{S\arabic{equation}}
\setcounter{theorem}{0}
\renewcommand{\thetheorem}{S\arabic{theorem}}

\section{Supplementary Information}\label{app}
\begin{subappendices}
\rev{
\subsection{Proof of Theorem~\ref{thm:strong-dob-average}}\label{app:sec:strong-dob-average}

Using Eqs.~\eqref{eq:bindex-strong-k} and \eqref{eq:contribution} one can derive 
Eq.~\eqref{eq:bindex-strong} as a weighted average of local balance:
\begin{equation}
\begin{split}
    R(\beta)
    &= \frac{\tr\mathbf{W}(\mathbf{A}, \beta)}{\tr\mathbf{W}(|\mathbf{A}|, \beta)} \\
    &= \frac{1}{\tr\mathbf{W}(|\mathbf{A}|, \beta)}
    \sum_k\frac{\beta^k}{k!}\tr\mathbf{A}^k \\
    &= \sum_k\frac{\beta^k}{k!} 
    \times \frac{\tr|\mathbf{A}|^k}{\tr\mathbf{W}(|\mathbf{A}|, \beta)}
    \times \frac{\tr\mathbf{A}^k}{\tr|\mathbf{A}|^k} \\
    &= \sum_kC_k(\beta)R_k
\end{split}
\end{equation}
Now, the above result can be rewritten in terms of Eqs. \eqref{eq:dob-strong} and \eqref{eq:dob-strong-k}:
\begin{equation}
\begin{split}
    R(\beta) 
    &= \sum_kC_k(\beta)R_k \\
    2B(\beta) - 1
    &= \sum_kC_k(\beta)(2B_k -1) \\
    &= 2\sum_kC_k(\beta)B_k - \overbrace{\sum_kC_k(\beta)}^{=1}
\end{split}
\end{equation}
The last term is equal to $1$ thanks to the normalization property of the contribution scores.
Thus, after some straightforward algebra, we have that:
\begin{equation}
    B(\beta) = \sum_kC_k(\beta)B_k
\end{equation}
which ends the proof. \qedsymbol
}

\newpage
\rev{
\subsection{Weak DoB as a weighted average}\label{app:sec:weak-dob-average}

\begin{theorem}\label{thm:weak-dob-average}
    Let $G$ be a signed graph, $\beta > 0$ a resolution parameter and 
    $2 \leq k = k_{min}, \ldots, k_{\max}$ a sequence of consecutive integers. Then:
    \begin{equation*}
        W(\beta) = \sum_{k}C_k(\beta)W_k
    \end{equation*}
\end{theorem}

\begin{proof}\label{proof:weak-dob-average}
We first rewrite Eq.~\eqref{eq:dob-weak} in terms of Eq.~\eqref{eq:weakly-unbalanced-walks-k}
and then follow with a few simple transformations to get the final result:
\begin{equation}    
\begin{split}
    W(\beta) 
    &= 1 - \frac{\tr\mathbf{V}(\mathbf{A}, \beta)}{\tr\mathbf{W}(|\mathbf{A}|, \beta)} \\
    &= 1 - \frac{\sum_k\frac{\beta^k}{k!}\tr\mathbf{V}_k(\mathbf{A})}{\tr\mathbf{W}(|\mathbf{A}|, \beta)} \\
    &= 1 - \sum_k\frac{\beta^k}{k!}\tr\mathbf{V}_k(\mathbf{A})\frac{1}{\tr\mathbf{W}(|\mathbf{A},\beta)} 
    \times \frac{\tr|\mathbf{A}|^k}{\tr|\mathbf{A}|^k} \\
    &= 1 - \sum_k\overbrace{\frac{\beta^k}{k!}\frac{\tr|\mathbf{A}|^k}{\tr\mathbf{W}(|\mathbf{A}|,\beta}}
    ^{C_k(\beta)}\frac{\tr\mathbf{V}_k(\mathbf{A})}{\tr|\mathbf{A}|^k} \\
    &= \sum_kC_k(\beta)\left(1 - \frac{\tr\mathbf{V}_k(\mathbf{A})}{\tr|\mathbf{A}|^k}\right) \\
    &= \sum_kC_k(\beta)W_k
\end{split}
\end{equation}
where in the second last equality we used the fact that $\sum_kC_k(\beta) = 1$.
\end{proof}
}

\rev{
\newpage
\subsection{Accuracy of numerical approximations}\label{app:sec:numerical}

MSB uses two different numerical approximations to attain high computational efficiency.
The first approximation happens when truncating power series to include only the terms of orders
$k_{\min}, \ldots, k_{\max}$. However, this approximation introduces no significant error by design,
as LP ensures that higher order terms have very low and monotonically decreasing contributions to the 
overall DoB calculations. Thus, as long as enough terms are included, 
and typically about a dozen or two is enough,
the truncation introduces no noticeable error. In principle, the lowest number of terms necessary
for attaining a given cumulative contribution score can be determined easily by inspecting the contribution
profile. However, here we used a simple rule-of-thumb and in all cases, unless specified otherwise,
used $k_{\max} = 30$, which is typically more than enough (see Fig.~\ref{app:fig:accuracy}A).

\begin{figure}[htb!]
\centering
\includegraphics[width=\textwidth]{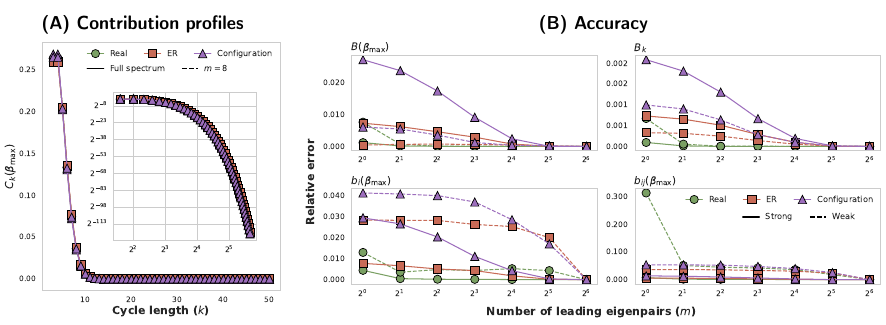}
\caption{
    Effects of MSB approximations assessed using bill co-sponsorship network
    from the U.S. Senate during 114th Congress ($|V| = 100$, $|E| = 3696$)
    as well as its randomized counterparts based on Erdős–Rényi model and configuration
    model~\cite{newmanNetworks2018}.
    \textbf{(A)}~Contribution profiles are clearly almost identical for the original and
    randomized networks. Crucially, at $\beta = \beta_{\max}$ almost all of the cumulative
    contribution score is driven by leading low order terms, meaning that higher order terms
    (roughly $k > 10$) can be safely omitted. The inset plot presents the same data on log-log scale
    in order to better display the tail of the distribution.
    \textbf{(B)}~Errors of DoB measures based on leading eigenvalues calculated relative
    to values obtained using full spectra. Errors are typically low even for $m = 1$
    and in all cases quickly decrease as $m$ increases. Moreover, in most of the cases
    they are lower for the real network 
    (as compared to its randomized counterparts based on the Erdős–Rényi and configuration models), 
    which is consistent with the fact that errors should be lower for networks with heterogeneous
    distributions of eigenvalues.
}
\label{app:fig:accuracy}
\end{figure}

The last approximation happens when only $m$ leading eigenpairs from the both ends of the spectrum
are used. This allows for solving the corresponding eigenproblems and running other downstream 
calculations much faster. Moreover, as discussed in Sec.~\ref{app:sec:analytic}, this approximation
is optimal and can be highly accurate, especially for real-world networks with heterogeneous spectra.
Fig.~\ref{app:fig:accuracy}B provides an empirical support for this claim.
}

\newpage
\subsection{Analytic functions of real symmetric matrices}\label{app:sec:analytic}

\rev{
Let $\mathbf{X} \in \mathbb{R}^{n \times n}$ be a real symmetric matrix and 
$f: \mathbb{R}^{n \times n} \to \mathbb{R}^{n \times n}$ an analytic function defined over
the field of real square matrices. 
Then, $f\left(\mathbf{X}\right) = \mathbf{Q}f\left(\mathbf{\Lambda}\right)\mathbf{Q}^\top$,
where $\mathbf{\Lambda}$ is a real diagonal matrix with eigenvalues of $\mathbf{X}$ 
satisfying $|\lambda_1| \geq |\lambda_2| \geq \ldots \geq |\lambda_n|$
and the columns of $\mathbf{Q}$ are the corresponding eigenvectors. 
This implies that:
\begin{align}
    f\left(\mathbf{X}\right)_{ij} &= \sum_{r=1}^n \mathbf{Q}_{ir}f(\lambda_r)\mathbf{Q}_{jr}
    \label{eq:methods:fXij}\\
    \tr{}f\left(\mathbf{X}\right) &= \sum_{r=1}^n f(\lambda_r)
    \label{eq:methods:trfX}
\end{align}
In particular, $k$th power is given by $\mathbf{X}^k = \mathbf{Q}\mathbf{\Lambda}^k\mathbf{Q}^\top$
and exponential by $e^{\mathbf{X}} = \mathbf{Q}e^{\mathbf{\Lambda}}\mathbf{Q}^\top$.

Note that Eqs. \eqref{eq:methods:fXij} and \eqref{eq:methods:trfX} can be approximated using only
$m$ leading eigenvalues, which allows for efficient computations.
In particular, as a consequence of Eckart-Young low-rank approximation 
theorem~\cite{eckartApproximationOneMatrix1936},
the error when reconstructing $\mathbf{Y} = f(\mathbf{X})$ based on $m$ leading eigenvalues and eigenvectors,
provided that $\mathbf{X}$ is symmetric and $f$ is analytic, is:
\begin{equation}\label{eq:methods:fX-error}
    \norm{\mathbf{Y} - \mathbf{\hat{Y}}}_F
    = \sqrt{\sum_{i=m+1}^n f(\lambda_i)^2}
\end{equation}
where $\mathbf{\hat{Y}}$ is the reconstructed matrix, and $\norm{\cdot}_F$ is Frobenius norm.
This approximation produces a matrix minimizing the error across all rank $m$ matrices and therefore
is optimal. Moreover, it is clear that if $|f(x)|$ is increasing the approximation is more accurate
for networks with heterogeneous spectra, or when some eigenvalues are much larger (in absolute value)
than others, which is a frequent property of real-world networks.
}

\newpage
\subsection{Computational complexity}\label{app:sec:efficiency}

\begin{figure}[htb!]
\centering
\includegraphics[width=.85\textwidth]{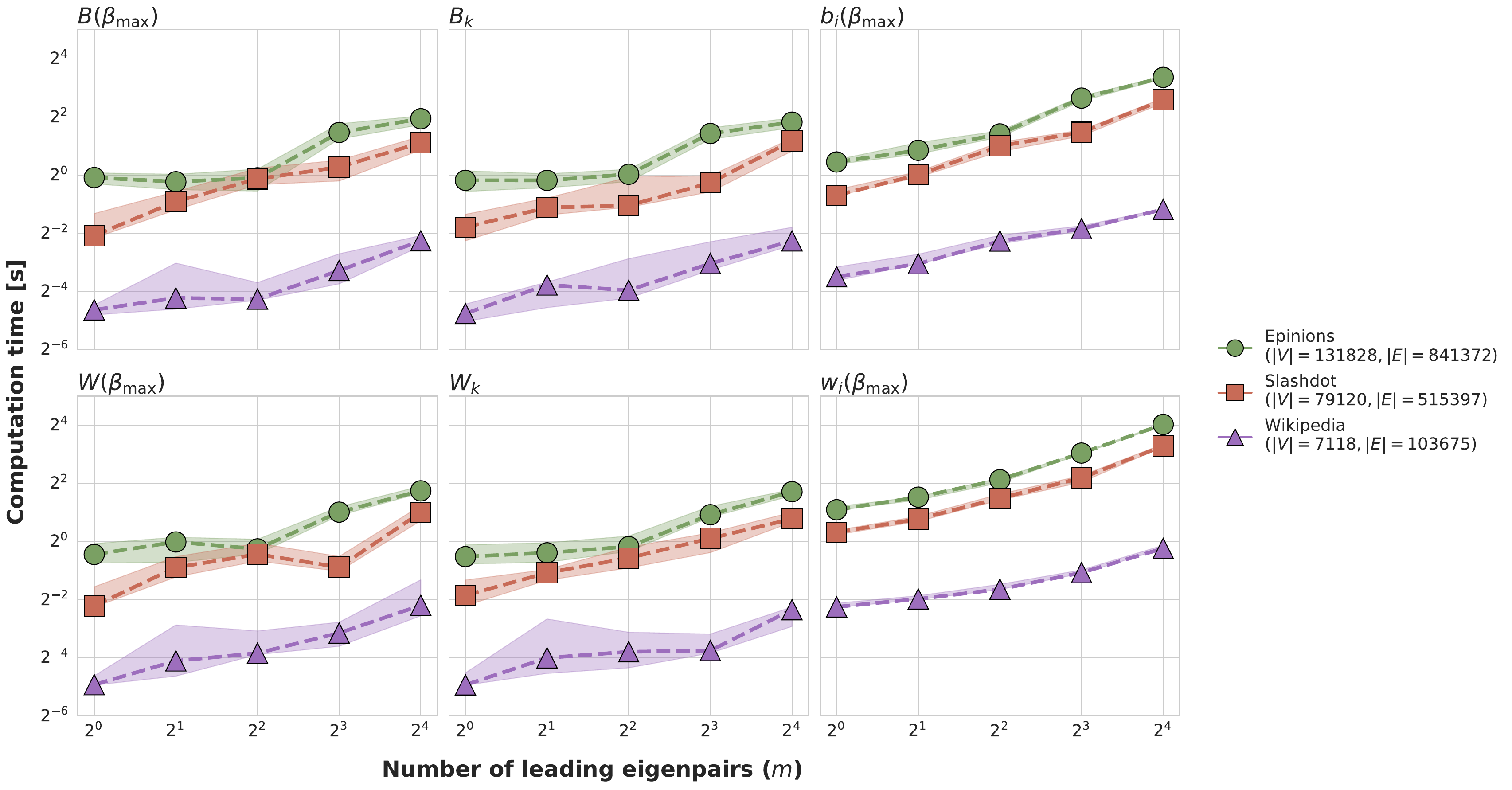}
\caption{
Running times of global, local and node-wise DoB measures (both strong and weak).
Lines correspond to median times (over 10 repetitions) and bounds to 1st and 9th deciles.
}
\label{app:fig:efficiency}
\end{figure}

Below are computation times for global, local and node-wise DoB measures for the three large
networks studied in this paper (Epinions, Slashdot and Wikipedia). Performance was assessed
using a laptop with AMD Ryzen 9 5900HX CPU and 32Gb of RAM. As evident in Fig.~\ref{app:fig:efficiency},
all running times were arguably short. Global DoB and balance profiles were calculated in about
1 second or much less. Node-wise measures (for all nodes) were calculated in no more than 
16 seconds (in the case of the largest network). All results include both the time needed for solving 
the eigenproblem(s), which can be cached and re-used in multiple computations, as well as any downstream
computations using eigenvalues and eigenvectors. Furthermore, in all cases computation times seem to 
scale with respect to $m$ in a very similar fashion with an average slope coefficient (in log-log scale)
of about $0.61$. This indicates that, at least for relatively low values of $m$, MSB computation times
are only moderately (sub-linearly) affected when increasing the number of used leading eigenpairs.

\newpage
\subsection{Descriptive statistics for the U.S. Congress co-sponsorship networks}
\label{app:sec:congress}

\begin{table}[ht!]
\footnotesize
\begin{threeparttable}
\begin{tabular}{ll|rrrrrrrrr}
\toprule
Chamber & Congress & $f_R$ & $f_D$ & $|V|$ & $|E|$ & $f_+$ & $B$ & $W$ & $\bar{d}$ & $d_{\text{cv}}$ \\
\midrule
\multirow[t]{21}{*}{House} & 93 & 0.43 & 0.56 & 446 & 18083 & 0.56 & 0.71 & 0.94 & 81.09 & 0.73 \\
 & 94 & 0.33 & 0.67 & 445 & 19503 & 0.58 & 0.71 & 0.95 & 87.65 & 0.70 \\
 & 95 & 0.34 & 0.66 & 444 & 21133 & 0.59 & 0.68 & 0.96 & 95.19 & 0.60 \\
 & 96 & 0.38 & 0.62 & 442 & 51081 & 0.84 & 0.55 & 1.00 & 231.14 & 0.33 \\
 & 97 & 0.45 & 0.54 & 447 & 49364 & 0.82 & 0.57 & 1.00 & 220.87 & 0.35 \\
 & 98 & 0.39 & 0.61 & 444 & 48721 & 0.75 & 0.59 & 0.99 & 219.46 & 0.31 \\
 & 99 & 0.42 & 0.57 & 443 & 49764 & 0.72 & 0.61 & 0.98 & 224.67 & 0.29 \\
 & 100 & 0.42 & 0.58 & 446 & 50688 & 0.73 & 0.62 & 0.99 & 227.30 & 0.29 \\
 & 101 & 0.42 & 0.58 & 449 & 56231 & 0.70 & 0.62 & 0.98 & 250.47 & 0.25 \\
 & 102 & 0.40 & 0.60 & 447 & 58067 & 0.68 & 0.63 & 0.98 & 259.81 & 0.25 \\
 & 103 & 0.42 & 0.58 & 446 & 59092 & 0.68 & 0.70 & 0.98 & 264.99 & 0.23 \\
 & 104 & 0.53 & 0.46 & 445 & 62154 & 0.72 & 0.78 & 1.00 & 279.34 & 0.25 \\
 & 105 & 0.52 & 0.48 & 449 & 66701 & 0.69 & 0.80 & 1.00 & 297.11 & 0.23 \\
 & 106 & 0.51 & 0.49 & 442 & 63652 & 0.67 & 0.83 & 0.99 & 288.02 & 0.24 \\
 & 107 & 0.51 & 0.48 & 447 & 63851 & 0.68 & 0.84 & 0.99 & 285.69 & 0.24 \\
 & 108 & 0.52 & 0.47 & 444 & 66277 & 0.67 & 0.84 & 0.99 & 298.55 & 0.24 \\
 & 109 & 0.53 & 0.47 & 445 & 66700 & 0.68 & 0.82 & 1.00 & 299.78 & 0.24 \\
 & 110 & 0.46 & 0.54 & 452 & 70923 & 0.67 & 0.82 & 0.99 & 313.82 & 0.20 \\
 & 111 & 0.41 & 0.59 & 451 & 70160 & 0.68 & 0.77 & 0.99 & 311.13 & 0.21 \\
 & 112 & 0.54 & 0.45 & 450 & 77872 & 0.66 & 0.82 & 1.00 & 346.10 & 0.18 \\
 & 113 & 0.53 & 0.47 & 447 & 75771 & 0.64 & 0.86 & 1.00 & 339.02 & 0.18 \\
 & 114 & 0.56 & 0.44 & 446 & 75180 & 0.65 & 0.86 & 1.00 & 337.13 & 0.19 \\
 \midrule
\multirow[t]{21}{*}{Senate} & 93 & 0.42 & 0.56 & 101 & 2439 & 0.76 & 0.62 & 0.99 & 48.30 & 0.35 \\
 & 94 & 0.37 & 0.61 & 100 & 2432 & 0.79 & 0.63 & 1.00 & 48.64 & 0.35 \\
 & 95 & 0.37 & 0.62 & 104 & 2336 & 0.81 & 0.57 & 1.00 & 44.92 & 0.39 \\
 & 96 & 0.41 & 0.58 & 101 & 2275 & 0.82 & 0.55 & 1.00 & 45.05 & 0.39 \\
 & 97 & 0.52 & 0.47 & 101 & 2073 & 0.79 & 0.60 & 0.99 & 41.05 & 0.37 \\
 & 98 & 0.53 & 0.47 & 101 & 2194 & 0.76 & 0.58 & 0.99 & 43.45 & 0.33 \\
 & 99 & 0.52 & 0.48 & 101 & 2177 & 0.75 & 0.61 & 0.99 & 43.11 & 0.33 \\
 & 100 & 0.46 & 0.54 & 101 & 2143 & 0.72 & 0.60 & 0.99 & 42.44 & 0.36 \\
 & 101 & 0.44 & 0.54 & 101 & 2445 & 0.68 & 0.63 & 0.98 & 48.42 & 0.31 \\
 & 102 & 0.42 & 0.56 & 102 & 2479 & 0.71 & 0.64 & 0.99 & 48.61 & 0.31 \\
 & 103 & 0.44 & 0.54 & 101 & 2257 & 0.72 & 0.70 & 0.99 & 44.69 & 0.35 \\
 & 104 & 0.52 & 0.46 & 102 & 2324 & 0.74 & 0.81 & 1.00 & 45.57 & 0.38 \\
 & 105 & 0.53 & 0.45 & 100 & 3002 & 0.70 & 0.81 & 1.00 & 60.04 & 0.27 \\
 & 106 & 0.53 & 0.45 & 102 & 2930 & 0.72 & 0.78 & 0.99 & 57.45 & 0.25 \\
 & 107 & 0.48 & 0.50 & 101 & 2522 & 0.73 & 0.70 & 0.99 & 49.94 & 0.30 \\
 & 108 & 0.50 & 0.48 & 100 & 2387 & 0.74 & 0.79 & 0.99 & 47.74 & 0.28 \\
 & 109 & 0.53 & 0.45 & 101 & 2823 & 0.73 & 0.82 & 0.99 & 55.90 & 0.25 \\
 & 110 & 0.49 & 0.49 & 102 & 2779 & 0.70 & 0.85 & 0.99 & 54.49 & 0.30 \\
 & 111 & 0.39 & 0.59 & 109 & 3645 & 0.74 & 0.68 & 1.00 & 66.88 & 0.27 \\
 & 112 & 0.48 & 0.50 & 101 & 3914 & 0.69 & 0.78 & 1.00 & 77.50 & 0.20 \\
 & 113 & 0.44 & 0.54 & 105 & 3932 & 0.65 & 0.86 & 1.00 & 74.90 & 0.24 \\
 & 114 & 0.54 & 0.44 & 100 & 3696 & 0.61 & 0.96 & 1.00 & 73.92 & 0.20 \\
 \bottomrule
\end{tabular}
\begin{tablenotes}
    \footnotesize
    \item $f_R, f_D$ -- fraction of Republicans/Democrats 
    (may not sum up to 1 due to the presence of other parties and/or independents)
    \item $|V|, |E|$ -- number of nodes/edges
    \item $f_+$ -- fraction of positive edges
    \item $B, W$ -- strong/weak degree of balance for $\beta \coloneqq \beta_{\max}$
    \item $\bar{d}, d_{\text{cv}}$ -- average degree and coefficient of variation of degree distribution
\end{tablenotes}
\end{threeparttable}
\end{table}

\end{subappendices}

\end{document}